\documentclass[twocolumn]{aa}
\usepackage{graphicx}
\usepackage{natbib}
\newcommand{\be}{\begin{equation}}
\newcommand{\ee}{\end{equation}}
\newcommand{\mytheta}{{\mbox{\boldmath$\vartheta$}}}
\newcommand{\mymag}{\hbox{$\,.\!\!^{\rm{m}}$}}
\newcommand{\myarcsec}{\hbox{$.\!\!^{\prime\prime}$}}
\newcommand{\myarcmin}{\hbox{$.\!\!^{\prime}$}}
\newcommand{\myarcsecnodot}{\hbox{$\;\!\!^{\prime\prime}\;$}}
\newcommand{\myarcminnodot}{\hbox{$\;\!\!^{\prime}\;$}}

\begin{document}
   \title{GaBoDS: The Garching-Bonn Deep Survey}

   \subtitle{I. Anatomy of galaxy clusters in the background of NGC 300\thanks{Based on 
observations made with ESO Telescopes at the La Silla Observatory}}
   \author{Mischa Schirmer\inst{1,2}, Thomas Erben\inst{1}, Peter Schneider\inst{1},
           Grzesiek Pietrzynski\inst{3,4}, Wolfgang Gieren\inst{3},\\
           Alberto Micol\inst{5}, Francesco Pierfederici\inst{6,5}
          }

   \offprints{M. Schirmer}

   \institute{Institut f\"ur Astrophysik und Extraterrestrische Forschung (IAEF), Universit\"at Bonn,
              Auf dem H\"ugel 71, 53121 Bonn, Germany; \email{mischa@astro.uni-bonn.de}
         \and
             Max-Planck-Institut f\"ur Astrophysik, Karl-Schwarzschild-Strasse 2, 85748 Garching, Germany
         \and
             Universidad de Concepci\'{o}n, Casilla 4009, Concepci\'{o}n, Chile
	 \and
             Warsaw University Observatory, Al. Ujazdowskie 4, 00-478 Warsaw, Poland
	 \and
             Space Telescope European Coordinating Facility, Karl-Schwarzschild-Strasse 1, 
             85748 Garching, Germany
	 \and
             National Optical Astronomy Observatory, 950 North Cherry Avenue, Tucson, USA
             }

   \date{Received ??.??, 2003; accepted ??.??, 2003}

   \abstract{The Garching-Bonn Deep Survey (GaBoDS) is a virtual 12 square degree cosmic shear and 
cluster lensing survey, conducted with the WFI@2.2m MPG/ESO telescope at La Silla. It consists of 
shallow, medium and deep random fields taken in $R$-band in subarcsecond seeing conditions at 
high galactic latitude. A substantial amount of the data was taken from the ESO archive, by 
means of a dedicated ASTROVIRTEL program.

In the present work we describe the main characteristics and scientific goals of GaBoDS. Our strategy 
for mining the ESO data archive is introduced, and we comment on the Wide Field Imager data reduction 
as well. In the second half of the paper we report on clusters of galaxies found in the background of 
NGC 300, a random archival field. We use weak gravitational lensing and the red cluster sequence 
method for the selection of these objects. Two of the clusters found were previously known and 
already confirmed by spectroscopy. Based on the available data we show that there is significant 
evidence for substructure in one of the clusters, and an increasing fraction of blue galaxies towards 
larger cluster radii. Two other mass peaks detected by our weak lensing technique coincide with red 
clumps of galaxies. We estimate their redshifts and masses. 

   }

\titlerunning{GaBoDS: Anatomy of galaxy clusters in the background of NGC 300}
\authorrunning{M. Schirmer et al.}

   \maketitle

\section{Introduction}
Light bundles from distant background sources are distorted by the tidal gravitational 
fields of the intervening matter distributed along the line of sight. This \textit{shear} can be 
statistically measured by determining the image ellipticities of those background sources. 
Since it is linearly related to the underlying matter distribution, the shear can be used 
to measure statistical properties of the large-scale structure in the Universe 
(\textit{cosmic shear}). It can also be used to detect clusters of galaxies, without assuming 
any hydrostatic equilibrium, symmetry or relation between luminous and dark matter. For recent 
reviews about weak gravitational lensing see \cite{mellier_review}, and \cite{bs_review}.

The amplitude of the distortions induced into galaxy images is of the order of a few percent 
for lensing by large-scale structure, and up to ten percent in the case of cluster lensing, 
depending on the mass of the cluster under consideration. The intrinsic ellipticities of galaxy
images dominate the noise of the shear signal. Thus, a weak lensing analysis requires deep 
data with good PSF properties in order to measure enough galaxy ellipticities with sufficient 
accuracy. A large field of view and independent lines of sight guarantee a representative view 
of the universe, beating down the effect of cosmic variance. Deep wide field surveys provide 
this kind of data.

The Garching-Bonn Deep Survey (hereafter: GaBoDS) is a mostly virtual survey and covers a sky area 
of more than $12$ square degrees with the Wide Field Imager\footnote{Hereafter we will use the term 
`WFI' for any Wide Field Imager instrument, and `WFI@2.2' when refering to the Wide Field Imager at 
the MPG/ESO 2.2m telescope.} at the MPG/ESO 2.2m telescope. Compared to other telescopes (VLT, CFHT, 
CTIO, WHT, KPNO) that have been successfully used for cosmic shear surveys so far \citep{bacon2000,
bacon2002,maoli,kaiser,waerbeke2000,waerbeke2001,wittman_cs}, the 2.2m telescope is small in size, 
yet it offers an outstanding image quality. It allows for an independent confirmation of previously 
obtained results, and the detection of possibly remaining systematics in the data as well as in the 
measurement process.

In this paper we use weak gravitational lensing to detect dark matter haloes by looking for coherent 
shear patterns in galaxy images \citep{schneider_map}. In this manner galaxy clusters are 
detected directly relying on their most fundamental property, their mass, which is independent of 
their dynamical state and luminosity. Since structure formation in the Universe is highly sensitive 
to the cosmological model, shear-selected samples of dark matter haloes will yield new insights 
into the process of cluster formation \citep[see][for shear-selected haloes]{erben_darkclump,margoniner,
miyazaki, dahle,wtm01,wtm02}. Based on simulations of the large-scale structure \citep{jain_waerbeke} 
as well as on predictions from Press-Schechter models \citep{kruse_pschechter} one expects some 
10 shear-selected dark matter haloes per square degree. Therefore, the number of mass peaks in the 
GaBoDS data is expected to be of the order of 100. The mass of these peaks strongly depends on their
redshift, as is shown in Sect. 5.

This work is organized as follows. In Sect. 2 we discuss our strategy for data mining the 
ESO archive. An outline of the main characteristics of the GaBoDS fields follows in Sect. 3,
together with a compact overview of our WFI data reduction process in Sect. 4.
Subsequently we report on the detection of galaxy clusters by means of weak lensing and the
red cluster sequence method \citep{gladder} in the NGC 300 GaBoDS field, and give a more 
detailed analysis for one of the clusters therein. We conclude in Sect. 6.
\section{Data mining the ESO archive}
\subsection{Our ASTROVIRTEL program}
For the above mentioned weak lensing analysis we aimed for a sky coverage of at least 10 square 
degrees for GaBoDS. Only about 3 square degrees were finally observed, however, in 20 allocated 
nights of our own GO program due to unfavourable weather conditions. Including available data 
from the EIS\footnote{ESO Imaging Survey} Deep Public Survey and COMBO-17\footnote{MPIA Heidelberg} 
left us with about 4 missing square degrees. A manual search in the large ESO archive turned out to 
be unfeasible, since the search engine available at that time did not allow for filtering the data 
with respect to our requirements. The only usable fields we knew beforehand in the archive were the 
ones from the Deep Public Survey and one pointing of the Capodimonte Deep Field. Other WFI@2.2 data 
such as the five COMBO-17 fields were taken during MPG time before the 2.2m telescope started into 
service mode operation, and were thus not publicly available through the archive. Besides, the very 
low number \footnote{Only about a dozen papers based on WFI@2.2 data have been published up to April 
2002.} of publications based on data taken with WFI@2.2 did not allow for a direct identification 
of further usable data. We therefore proposed an ASTROVIRTEL program\footnote{ASTROVIRTEL cycle 2: Erben 
et al., \textit{ Gravitational lensing studies in randomly distributed, high galactic latitude fields}}
\footnote{http://www.stecf.org/astrovirtel}, aiming at an enhancement of the \textit{querator}
\footnote{http://archive.eso.org/querator} search engine's capabilities \citep{pierfederici}.
In order for a field to be included in the GaBoDS, the following requirements addressed by 
\textit{querator} had to be met:
\begin{itemize}
\item{minimum exposure time in $R$-band: $\sim 5$ ksec,}
\item{\textit{image} seeing $\leq 1\myarcsec 0$,}
\item{random, at high galactic latitude,}
\item{\textit{empty}, i.e. avoidance of known very massive structures in or next to 
the field, no bright stars or large foreground objects inside the field.}
\end{itemize}
The first item in this list guarantees a high enough number density ($\geq 10$ arcmin$^{-2}$)
of galaxies with securely measurable shapes. Furthermore, exposures must be taken in excellent 
seeing conditions, since the S/N for shape measurements decreases with the second power of
the size of the PSF. The random character of the fields is needed to avoid a biasing towards 
certain types of objects, such as quasars or clusters of galaxies. In addition, it is ensured 
in this way that the fields sparsely sample the Universe along independent lines of sight,
thus keeping the effect of cosmic variance small. The last point in the above item list is to 
further guarantee that a given pointing is usable for our purposes: massive structures such as
large clusters of galaxies bias the search for unknown dark matter haloes as well as the 
measurement of a cosmic shear signal, which is about one order of magnitude smaller than 
the weak lensing signal of a large cluster. We are aware of the fact that this introduces a
bias towards lower density lines of sight. By avoiding bright and large foreground objects 
we keep the usable area of an image as large as possible.
\subsection{About data quality}
It is clear that `data quality' is a very ambiguous term, highly dependent on the science which is 
to be drawn from the data. Given the involved and time consuming reduction of WFI data, we wanted 
to assess as best as possible the quality of archival data before any request or data reduction. 
This is straight forward for items such as the total exposure time, available filters, presence of 
bright objects, availability of calibration frames, and the ambient conditions during which the 
observation were performed (moon; clouds; seeing, according to the seeing monitor). However, not 
all of those can be expressed in terms of numbers; some judgement has to be done upon visual 
inspection of the data. Other data quality issues, such as image seeing or PSF properties can not be 
evaluated without more complex operations on the data itself. To address most of these points, the 
following demands were defined to ASTROVIRTEL and \textit{querator}:
\begin{itemize}
\item{measure the image seeing for all archived WFI@2.2 data,}
\item{provide a preview facility, allowing quick visual inspection of the data ahead 
of a request,}
\item{make available the according proposal abstract (`why was this particular observation done?')}
\end{itemize}
\begin{table}[t]
   \caption{\label{astrovirtel_fields}Fields identified by our ASTROVIRTEL program. The 
second column gives the exposure time that went into the coaddition (images with very 
bad PSF or seeing were rejected). Exposure times in parentheses indicate the total 
exposure time per association as identified by \textit{querator}.}
   \begin{tabular}{lll}
   \hline 
   \noalign{\smallskip}
   Target & Exposure time & Image seeing\\
   \noalign{\smallskip}
   \hline 
   \noalign{\smallskip}
   B08p3    &   4900 (5600) &   0\myarcsec88\\
   B08p2    &   4900 (5600) &   0\myarcsec88\\
   B08p1    &   4500 (5600) &   0\myarcsec88\\
   B0800    &   7200 (8300) &   0\myarcsec88\\
   B08m1    &   4500 (6000) &   0\myarcsec88\\
   B08m2    &   4000 (5800) &   0\myarcsec88\\
   B08m3    &   5400 (6000) &   0\myarcsec96\\
   Pal3    &    5000 (6120) &   1\myarcsec0\\
   AM1     &    7500 (7620) &   1\myarcsec0\\
   Comparison1 & 5300 (9300) &  0\myarcsec97\\
   C04p1    &   4000 (5600) &   0\myarcsec88\\
   C04p3    &   4000 (5600) &   0\myarcsec83\\
   C04m2    &   4000 (4800) &   0\myarcsec85\\
   C04m3    &   4000 (4800) &   0\myarcsec85\\
   NGC300    &   15100 ($\approx 25000$) & 1\myarcsec06\\
   C04p2    &   4000 (5600) &  0\myarcsec86\\
   C04m1    &   4000 (4800) &  0\myarcsec86\\
   C04m4    &   4000 (4800) &  0\myarcsec86\\
   C0400    &   4800 (5600) &  0\myarcsec87\\
   \noalign{\smallskip}
   \hline 
   \end{tabular}
\end{table}
All but the last item were implemented. The image seeing, crucial for a weak lensing analysis, 
was determined by automatically extracting all non-saturated stars from one of the eight WFI@2.2 
chips, and averaging their FWHM. It was superior to the DIMM\footnote{Difference Image Motion 
Monitor} seeing, since the DIMM on La Silla could not pick up local effects such as dome seeing,
zenith distance, flexure and temperature of the telescope's Serrurier structure, focus and tracking.
On average we found the DIMM seeing to be $0\myarcsec1-0\myarcsec2$ better than the image seeing for 
WFI@2.2. It was only in rare cases that the difference between the two drops below $0\myarcsec1$.

In this way a list of useful candidate fields was extracted from the archive, minimizing the amount
of unusable data slipping into the reduction process. Fields that were rejected at this late 
step suffered from scattered light or had very bad PSF properties\footnote{The anisotropy threshold 
depends on the overall quality of a particular data set. In general we rejected exposures with 
anisotropies larger than $\sim 6\%$.}. Checking for PSF anisotropies in an image was very time 
consuming and thus not blindly performed on all data in the archive.

Using the enhanced \textit{querator} we found about $5$ square degrees ($20$ pointings) of data in 
the ESO archive which passed our criteria, not counting the already known fields such as the ones 
from the EIS Deep Public Survey. This was twice the area we expected, but it must be 
noted that 15 of the pointings were done by a single observer, searching for trans-neptunian objects. 
The efficiency of our archive search with querator was about $75\%$, i.e. three out of four candidate 
fields were usable. Table \ref{astrovirtel_fields} lists the fields found together with their image 
seeing.
\section{Main characteristics of the GaBoDS fields}
\begin{figure}[t]
  \includegraphics[width=1.0\hsize]{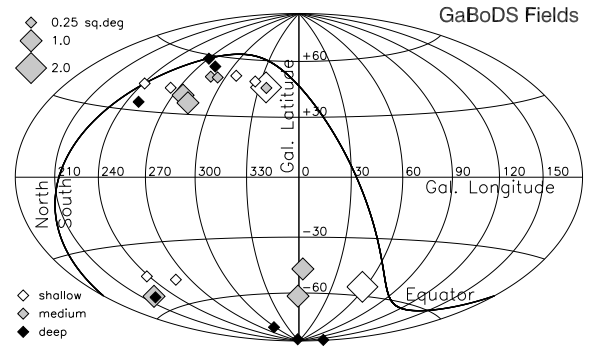} 
  \caption{\label{allsky}Sky distribution of the GaBoDS fields. The size of the symbols depicts
the covered sky area. All fields are at high galactic latitude.} 
\end{figure}

\begin{figure}[t]
  \includegraphics[width=1.0\hsize]{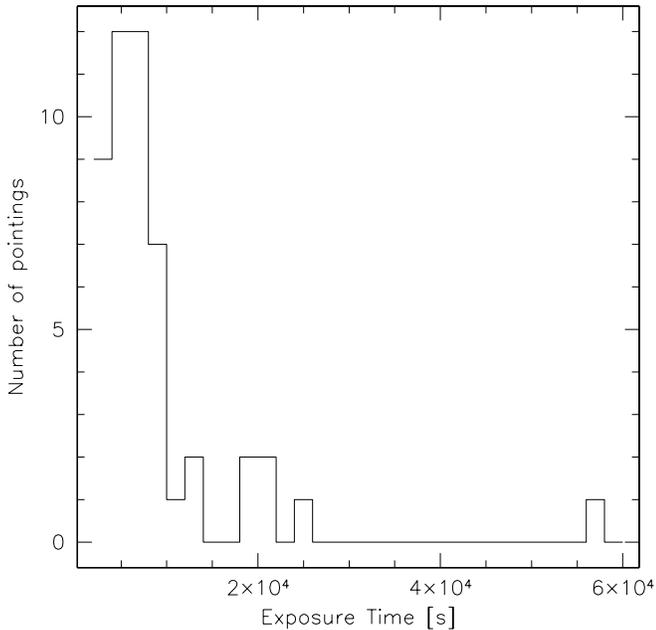}
  \caption{\label{time}Exposure times in GaBoDS. The peak at 56 ksec represents the Chandra 
Deep Field South (CDF-S).}
\end{figure}

Since we made heavy use of archival data, we were able to collect more than 12 square degrees of high 
quality $R$-band images with only 10 clear nights of own observations. For nearly half of the fields 
multicolour information is also available. As can be seen from Figs. \ref{allsky} and \ref{time}, the 
GaBoDS fields can be split into a shallow part (6 square degrees, 4-7 ksec total exposure time), a 
medium deep part (4.25 square degrees, 8-11 ksec) and a deep part (2.0 square degrees, 13-56 ksec). 
The survey fields are randomly distributed at high galactic latitude in the southern sky. Data in 
GaBoDS that was not taken by ourselves is summarised in Table \ref{GaBoDS_data}.

\setcounter{topnumber}{1}
The image seeing in the stacked images is equal to or better than 1\myarcsec0, and our astrometric
solution is accurate enough not to introduce artificial PSF anisotropies (see Sect. 4 for more details). 
The typical PSF quality of a stacked set of WFI@2.2 exposures in our survey can be seen in the lower 
right panel of Fig. \ref{psf_overview}. There we show the PSF anisotropies of a coadded image consisting 
of 57 exposures, giving a total integration time of 27.1 ksec. The image seeing of this particular field 
is $0\myarcsec8$, thus possible shortcomings in the astrometric algorithm could easily be seen. The rms 
PSF anisotropy merely amounts to $0.8\%$, making WFI@2.2 very well suitable for weak gravitational lensing 
studies. The larger anisotropies in the field corners, especially in the lower left, are due to slightly 
tilted CCDs with respect to the focal plane (K. Meisenheimer, private communication). Optical abberations 
play only a minor role for this instrument.

Such quality of the data can only be achieved with a very carefully and frequently refocused telescope.
The effect of a slightly defocused telescope on the PSF is shown in the remaining three panels of Fig. 
\ref{psf_overview}. One can see that anisotropies become significant once the detector is out of focus.
Furthermore, the PSF rotates by 90 degrees when one passes through the focal plane. This is characteristic 
for tangential and sagittal astigmatism. Still the PSF of WFI@2.2 is very homogeneous over the field of 
view, even when crossing chip borders. Thus larger dither patterns can be used for the observations, and 
a single smooth model can be fitted to the PSF in the stacked mosaic (see also Fig. \ref{ngc300_aniso}
in Sect. 5).
\begin{table}[t]
   \caption{\label{GaBoDS_data}GaBoDS data that was not taken by ourselves.} 
   \begin{tabular}{ll}
   \hline 
   \noalign{\smallskip}
   Sky coverage & Data source, depth of field\\
   \noalign{\smallskip}
   \hline 
   \noalign{\smallskip}
      1.75 $\mathrm{deg}^2$ & ESO Imaging Survey + GOODS\\
                   & (medium, one deep)\\
      5.00 $\mathrm{deg}^2$ & ASTROVIRTEL (mostly shallow, one deep)\\
      1.25 $\mathrm{deg}^2$ & COMBO-17 survey (deep)\\
      1.25 $\mathrm{deg}^2$ & ESO distant cluster survey\\
                   & (shallow, medium)\\
      0.25 $\mathrm{deg}^2$ & NTT Deep Field, from the IR group at the\\
                   & MPI f\"ur extraterrestrische Physik (deep)\\
      0.25 $\mathrm{deg}^2$ & Capodimonte Deep Field (deep)\\
   \noalign{\smallskip}
   \hline 
   \end{tabular}
\end{table}
\section{The art of WFI data reduction}
\subsection{The GaBoDS pipeline}
\setcounter{totalnumber}{4}
\setcounter{topnumber}{2}
\renewcommand{\textfraction}{0.05}
\renewcommand{\topfraction}{0.95}
The advent of multichip CCD cameras imposes new, high demands on data reduction. Pre-processing 
steps such as debiasing or flatfielding can be done independently on a chip-by-chip basis, 
allowing for efficient parallel processing on a multi-processor machine with sufficient disk 
space. Whereas these steps can be tackled using the same well-known algorithms as for single 
chip cameras, an accurate astrometric and photometric calibration of WFI data requires 
techniques going well beyond those routines. Different sensitivities of the CCDs and gaps 
between them lead to a very inhomogeneous exposure time and accordingly noise in the coaddition. 
An accurate weighting scheme is essential in order to retain control over these effects in 
the stacked image. In the following we describe our approach to WFI data reduction.

An almost fully automatic pipeline for WFI reduction was developed, based on existing software 
modules wherever possible, such as \textit{EIS drizzle}\footnote{http://www.eso.org/science/eis}, 
the Terapix\footnote{http://terapix.iap.fr} software suite, \textit{FLIPS}
\footnote{http://www.cfht.hawaii.edu/$\sim$jcc/Flips/flips.html} (J.-C. Cuillandre, not yet publicly 
available), \textit{Eclipse}\footnote{http://www.eso.org/projects/aot/eclipse}, \textit{Imcat}
\footnote{http://www.ifa.hawaii.edu/$\sim$kaiser/imcat} (N. Kaiser) and \textit{LDAC}
\footnote{ftp://ftp.strw.leidenuniv.nl/pub/ldac/software} (E. Deul, Leiden Data Analysis Center) 
(catalogue format and handling). \textit{IRAF} was not used in the pipeline, apart from the drizzle 
coaddition, since it did not allow for efficient scripting and reduction of this specific kind of data. 
A number of bash shell scripts were wrapped around those tools (mostly stand-alone C programs), 
allowing for an efficient, flexible and almost fully automatized end-to-end reduction of WFI 
data in parallel mode. The usage of the pipeline is not restricted to the WFI@2.2, but data from
other instruments such as FORS1/2@VLT, ISAAC@VLT, SUPRIMECAM@SUBARU, MOSAIC-I@CTIO, MOSAIC-II@KPNO and 
WFI@AAO has already been successfully reduced. Supported architectures are Solaris, AIX, Linux and 
Dec-Alpha. The package will be released together with a detailed technical description (Erben et al., 
in preparation).

Our pipeline was designed with the GaBoDS data in mind, i.e. empty fields at high galactic latitude, 
with a fairly large dither pattern of up to 3\myarcmin0, and a very large number of single exposures 
per pointing. Still, the usage of the GaBoDS pipeline is not resticted to empty fields only. This is 
shown in Sect. 5, where we present an analysis of galaxy clusters behind NGC 300. 
\begin{figure*}[t]
  \includegraphics[width=1.0\hsize]{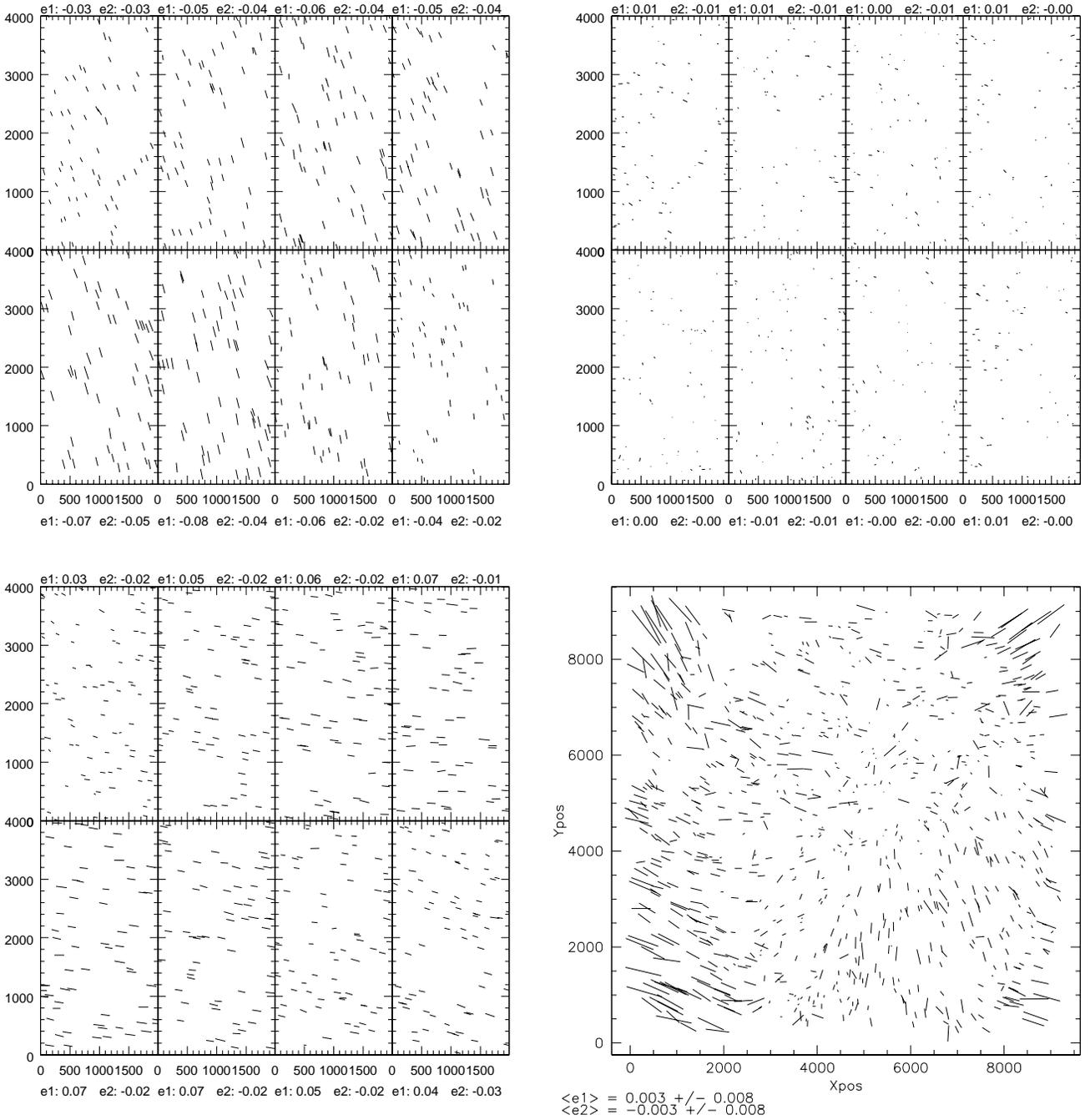}
  \caption{\label{psf_overview}PSF anisotropies for an intrafocal (upper left), focal (upper right) 
and extrafocal (lower left) exposure. The chosen scale for the stick length is the same for those 
three plots in order to show the increase in the anisotropies with respect to the focused exposure. 
The mean stellar ellipticities are 6.6\%, 0.9\% and 5.9\%, respectively. The lower right panel depicts 
typical PSF anisotropies of a stacked WFI@2.2 $R$-band image ($\sim50$ exposures with $\sim500$
sec exposure time each). Note that the largest PSF anisotropy in the stacked image is as small as 
0.8\%. Compared to the other three PSF plots a different scale for the stick length was used
in order to clearly show the anisotropies.}
\end{figure*}
\subsection{The pre-processing}
{\bf Overscan correction, debiasing, flat fielding:}
Apart from the astrometric, photometric and coaddition processes, all chips are processed 
individually, allowing for an easy parallelization of the code.

During pre-reduction, any instrumental signatures present in the data are removed. This includes 
overscan correction, bias subtraction and flat fielding with skyflats\footnote{For WFI@2.2m 
telescope dome flats are inferior compared to skyflats.}. For the master biases all bias frames 
are median combined with outlier rejection. Flat fields are combined in the same way, but each 
flat exposure is normalised to 1 before the combination. Thus the different gains in the science 
images are still present after the flat fielding step. This is because we found that the chip-to-chip 
gain variations can be better determined from a superflat, which is drawn from already flatfielded 
exposures.

{\bf Creation of a superflat:}
Residuals of around 3-4\% after normal flat fielding are common for WFI@2.2m,
depending on sky brightness and the filter in use. A superflat is computed for correction of this 
effect, by median combining all images from a given observing run, using outlier rejection. Pixels 
that are affected by stars or galaxies are detected with \textit{SExtractor} \citep{sextractor} and masked 
beforehand. In doing so we prevent the bright extended haloes around stars contributing to the superflat. 
The image constructed this way is then heavily smoothed, yielding an illumination correction image for 
every chip. All images are divided by their individual illumination correction. They are also normalised 
to the same (the highest) gain, which is accurately determined by comparing the modes of the individual 
superflats. Remaining residuals in the sky background are typically below 2\%, and in the case of 
absence of bright stars, even below 1\%.

{\bf Defringing:}
Besides the illumination correction a fringing model is calculated by subtracting the illumination 
correction from the previous superflat. Hereby it is assumed\footnote{This assumption only holds for 
photometric nights. In non-photometric nights it can be impossible to remove fringing in the $I$-band, 
since the model is changing within minutes on scales smaller than the field of view of a single CCD.} 
that the amplitudes of the observed fringes scale directly with the sky background. The 
fringing model is then individually scaled for each image and subtracted. Fringing in the $R$-band 
is of the order of a few percent for WFI@2.2. Its correction makes sense only if more than $\sim10$ 
images were used in the calculation of the model. Otherwise the pixel-to-pixel noise in the fringing 
model is larger than the fringing amplitude itself. This would introduce more noise into the individual 
defringed images than what is taken out by the correction of the lower frequency fringes. Since we are 
interested in measuring shapes of faint and small galaxies we want to avoid additional pixel-to-pixel 
noise. However, all images in the GaBoDS survey were constructed from many more than just 10 images, 
thus the contribution of additional high frequency noise is small, and we profit from taking out the 
gentle fringing pattern. In the case of the redder $I$-band the fringing can be much more prominent, 
and is in general more difficult to remove.
\subsection{Astrometric calibration}
After the pre-processing a global astrometric solution and a global relative photometric solution 
is calculated for all images. This is where the reduction of WFI data becomes much more complicated 
than the one for single chip cameras.

\begin{figure}[t]
  \includegraphics[width=1.0\hsize]{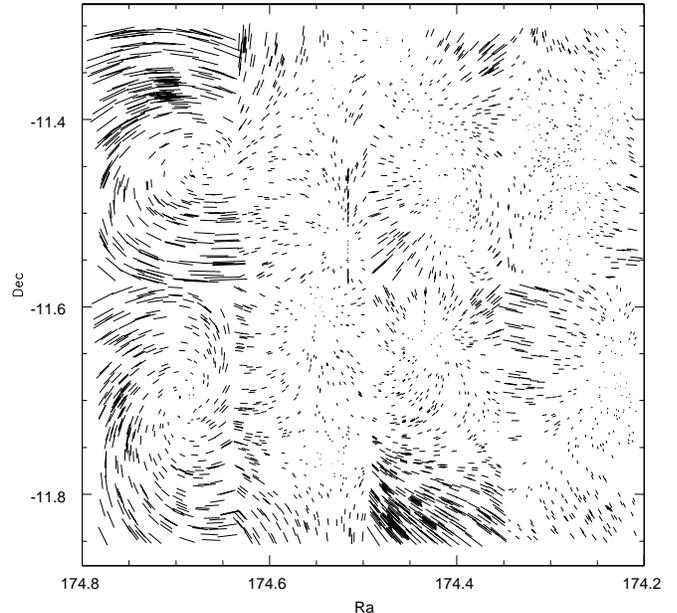}
  \caption{\label{WFI_distort}Difference in object position between a single-shift approach and a 
full two-dimensional second order astrometric solution for the WFI@2.2. In other 
words, shown are the higher order terms needed for matching the CCDs to the sky. The patterns belonging
to the left two chips are due to a rotation with respect to the mosaic. The maximum position difference 
in the plot is about six pixels, still a fairly small value compared to other telescope designs.
It becomes clear that a single, global distortion polynomial for all CCDs does not work. Instead, 
every CCD has to be treated individually.}
\end{figure}

In the first step, high S/N objects in each image are detected by \textit{SExtractor}, and a catalogue of 
non-saturated stars is generated. Based on a comparison with the USNO-A2 astrometric reference catalogue, 
a zero-order, single shift astrometric solution is calculated for each chip in every exposure. For a 
single-chip camera with a small field of view such an approach is often sufficient, but it no longer 
holds for multiple chip cameras with a large field of view. CCDs can be rotated with respect to each 
other, tilted against the focal plane, and in general cover areas at a distance from the 
optical axis, where field distortions become prominent\footnote{A ZEMAX analysis (Philipp Keller, 
private communication) for the 2.2m Ritchey Chretien telescope design and WFI@2.2's focal reducer 
shows that the radial field distortion for this layout increases with $\delta=a_1 r^2 + a_2 r^6$. 
However, the total amplitude of this distortion is very small ($\sim30$ pixels).}. Fig. 
\ref{WFI_distort} shows the difference between a zero order (single shift with respect to a reference 
catalogue) and a full astrometric second order solution per CCD. From this figure it is obvious that 
the simple shift-and-add approach will not work for the whole mosaic. The issue is further complicated 
by the gaps between the CCDs and large dither patterns that are used to cover them. Thus, chips with 
very different distortions overlap. In addition, due to the large field of view, the observed patch of 
the sky must no longer be treated as a flat plane, but as a spherical curved surface. 

In the second step we use Mario Radovich's \textit{Astrometrix}
\footnote{http://www.na.astro.it/$\sim$radovich/WIFIX/astrom.ps} (Terapix) package to fit third order 
polynomials to every chip in every exposure, in order to correct for the above mentioned effects and to
find a global astrometric solution. For this purpose all high S/N objects (stars and galaxies) detected 
in the first step are identified with each 
other, including those from the overlap regions. The latter ones are most important in establishing a 
global astrometric (and photometric) solution, since the accuracy of available reference catalogues 
such as the USNO-A2 (0\myarcsec2 rms) is insufficient for sub-pixel registration. Thus the astrometric 
solution is determined from the data itself. The USNO-A2 is used only to fix the solution with respect 
to absolute sky coordinates within 0\myarcsec2 rms. With \textit{Astrometrix} we consistently achieve 
an internal astrometric accuracy of $1/20-1/10$th of a pixel (0\myarcsec02--0\myarcsec01), thus the 
final PSF is mostly determined by the intrinsic PSFs of the single exposures (see Fig. \ref{psf_overview}). 
Additional, artificial seeing and PSF anisotropies are introduced into the stacked image on a very low 
level only, even for very large data sets such as the CDF-S, consisting of 150 WFI@2.2 R-band exposures.
This is a crucial requirement for our weak lensing analysis.
\subsection{Photometric calibration}
Once an astrometric solution is found, a relative photometric solution is straight forward. 
Relative fluxes of objects in different exposures and overlap regions are compared, allowing the 
calculation of relative photometric zeropoints for every chip and every exposure. Given two 
overlapping chips $k$ and $j$, consider all $i=1...N$ objects and calculate the mean deviation 
of magnitudes $K$ and $J$
\begin{equation}M_{k,j}:=\frac{\Sigma_{i}W_{i}(K_{i}-J_{i})}{\Sigma_{i}W_{i}}\;,
\end{equation}
with $W_{i}=(\sigma^{2}_{K}+\sigma^{2}_{J})^{-1}$, where ${\sigma}$ are the measurement errors of 
the corresponding magnitudes. Objects deviating in $K_{i}-J_{i}$ more than a user defined threshold 
are rejected. The relative zeropoints $ZP_{l}$ for all $N_{\mathrm{over}}$ overlapping CCDs are 
determined by $\chi^{2}$ minimization with respect to $ZP_{k}$:
\begin{equation}
\chi^{2}=\sum_{k,j}^{N_{\mathrm{over}}}\left[M_{k,j}-(ZP_{k}-ZP_{j})\right]^{2}\;.
\end{equation}
Finally, the relative zeropoints of photometric images are normalised so that their mean is zero.
This approach assumes that the relative zeropoints are constant for every CCD\footnote{Zeropoint 
variations for images taken with WFI@2.2 that were not superflat corrected are described in:\\
http://www.ls.eso.org/lasilla/sciops/wfi/zeropoints .}. An automatic absolute photometric solution 
is not yet implemented in the pipeline.
\begin{figure}[t]
  \includegraphics[width=1.0\hsize]{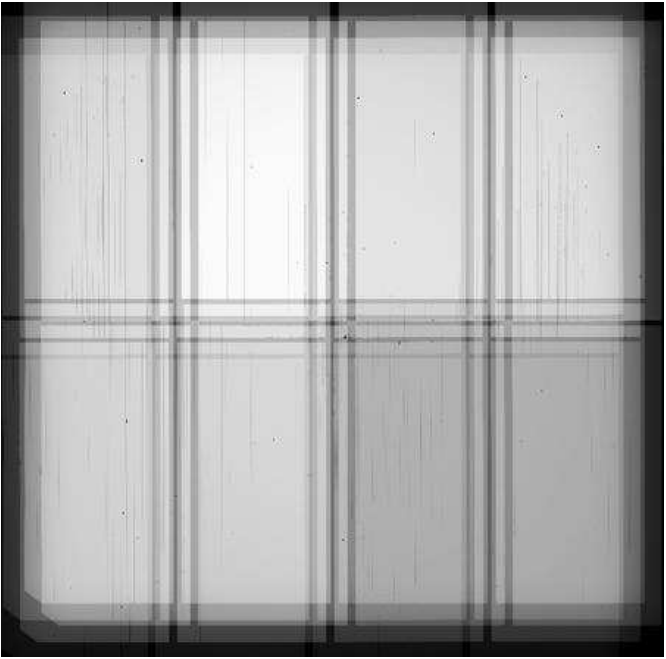}
  \caption{\label{weight_greyscale}Coadded weight image of a small WFI@2.2 data set consisting of five 
exposures. One clearly identifies regions with less effective exposure time due to gaps between 
CCDs and different pixel sensitivities. The size of the dither pattern also becomes obvious.
Brighter regions correspond to pixels with higher weight. The variations from chip to chip are due
to differences in the gain and the flatfield.}
\end{figure}
\subsection{A statistically optimized weighting scheme}
The effective exposure time for a stacked WFI mosaic is highly non-uniform. Read noise and the 
flatfields are chip-dependent, and gaps between the CCDs contribute further to the 
inhomogeneous depth of a stacked image. Applying a statistically optimized weighting scheme 
to the exposures during coaddition allows for a significantly improved object S/N ratio 
(up to a factor of $\sim1.5$).

We now describe our approach to assign an individual weight map for every science image.
In a first step, a pixel is assigned its normalised skyflat value as a weight, which contains the 
information about relative gains between the CCDs and pixel-to-pixel sensitivities. Contrary to 
other methods, we do not detect `bad' pixels (hot or cold pixels, pixels affected by cosmics, 
reflections or satellite tracks) by intercomparing all images in the stack, but on the individual 
images themselves. For the detection of permanent image defects, such as hot or dead pixels and 
bad columns, we use dark frames and superflat images. Affected pixels are set to zero in the 
corresponding weight map. Thus every chip in the WFI mosaic has its own basic weight map after the 
first step. Weight maps created in this way are the same for all exposures in a data set unless 
they were flatfielded with different skyflats or taken several weeks apart.

In a second step these weight images are adjusted individually for every image. Remaining hot pixels 
and cosmics are easily identified with \textit{SExtractor} in conjunction with \textit{Eye} (Terapix), 
since they appear much sharper than the stellar PSF even under good seeing conditions. Bright reflections 
and satellite tracks, however, need to be masked by hand, the only step in the pipeline which is not 
yet automatized. Moving objects like asteroids go unmasked and show up as dashed lines in the stacked 
image. During coaddition the individual weight maps are scaled with correction factors for airmass 
and varying photometric conditions. Changing seeing conditions from image to image can be 
included on an optional basis, too. All individual weight maps are resampled and coadded
in exactly the same way as the respective science images, yielding the noise properties for 
all pixels in the final coadded image (see Fig. \ref{weight_greyscale} for an example).
\subsection{The coaddition process}
Before the coaddition, all images are sky subtracted. In order to model the sky 
background we detect all objects in the field with \textit{SExtractor} and replace them with the 
mean background as determined from the remaining pixels. This image is then convolved 
with a very broad smoothing kernel (width between 200 and 500 pixels) and subtracted 
from the science image itself.

For the coaddition the \textit{EIS drizzle} in \textit{IRAF} is used. It allows for a weighted mean 
coaddition, guaranteeing the best S/N in the stacked image. The resampling strategy ensures 
that the PSF is not artificially bloated in the stack, and that a varying pixel scale is 
correctly taken into account also from a photometric point of view. Alternatively, one can use
\textit{swarp} (E. Bertin, Terapix), which contrary to \textit{drizzle}, makes use of much more 
advanced resampling algorithms and does not lead to correlated noise in the stacked image. 
However, the differences between \textit{swarp} and \textit{drizzle} vanish if many images are 
stacked, which is the case in GaBoDS.

Four factors determine the value of an output pixel in the coadded image. 
We have the input pixel value $I_i$ from the science chip and an associated value $W_{i}$ in 
the weight map. $I_i$ represents the part of the input pixel that is mapped onto the 
corresponding output pixel. Besides, $I_{i}$ is scaled with factors $f_{i}$ to the consistent 
photometric zeropoint and to a fixed exposure time (we chose 1 sec for this purpose):
\begin{equation}
f_{i}=10^{-0.4\,ZP_{i}}/t_{i},
\end{equation}
where $t_i$ is the exposure time and $ZP_i$ the relative photometric zeropoint. All images 
are also weighted according to their sky noise. This weight scale is given by:
\begin{equation}
w_i=\frac 1{\sigma_{\rm sky,i}^2 f_i^2}.
\end{equation}
Here we take into account the fact that the noise also scales with the flux scale $f_i$. 
The values $I_{\rm out}$ and $W_{\rm out}$ in a stack of $N$ images then read
\begin{eqnarray}
I_{\rm out}=\frac{\sum_{i=1}^{N}I_if_iW_iw_i}{\sum_{i=1}^{N}W_iw_i} & , \;\;\;\;\;&
W_{\rm out}=\sum_{i=1}^{N}W_iw_i\;.
\end{eqnarray}
\textit{EIS drizzle} creates its output with the TAN projection. Alternatively, the COE projection 
can be used \citep[see][for further information on sky projections]{greisen}. In the stacked images 
North is up and East to the left. A reference coordinate can be specified for the coaddition. Thus, 
if multicolour information is available for a particular pointing, the stacked images in the different 
bands are automatically registered with subpixel accuracy.
\section{Galaxy clusters in the background of NGC 300}
\subsection{Characteristics of the data}
One of the ASTROVIRTEL fields was a deep multicolour observation of NGC 300 (see Table 
\ref{astrovirtel_fields}), a face-on spiral galaxy in the Sculptor group at a distance of 
about 2.1 Mpc \citep{freedman}. Its angular size is $25\myarcminnodot\times18\myarcminnodot$, 
occupying about 40\% of the WFI@2.2 field of view ($34\myarcminnodot\times34\myarcminnodot$). The field 
did thus not meet all requirements for GaBoDS, but the image seeing of a significant fraction of the 
$R$-band data was around 1\myarcsec0. Deep $V$-band observations ($\sim37$ ksec, 1\myarcsec13 image 
seeing) were available too. This allowed us to search for distant galaxy clusters in the background 
of NGC 300 by means of weak lensing and the red cluster sequence method.

Upon visual inspection in the Digitized Sky Survey (DSS) before the data retrieval request, we 
recognized two concentrations of fainter galaxies north-east and south-east of NGC 300, at the edge of 
the WFI@2.2 field. 21 spectra were taken for the first concentration by \citet{cappi} (hereafter CHM98), 
confirming a cluster at redshift 0.165. The second, less prominent concentration is known as EDCC-499 
at a redshift of 0.117 \citep{collins}. Hereafter we refer to these clusters as CL0056.03--37.55 
(CL0053--37 in CHM98) and CL0056.02--37.90, respectively, using their epoch 2000.0 equatorial coordinates. 
The WFI@2.2 data for NGC 300 was taken for the ARAUCARIA project \citep{pietrzynski}, an attempt to 
fine-tune the distance ladder by comparing different distance indicators such as cepheids, blue 
supergiants, tip of the red giant branch, and planetary nebulae for various nearby galaxies. Since 
NGC 300 was monitored in 34 nights between July 1999 and January 2000, the creation of deep 
multicolour images was feasible.
\begin{figure*}[t]
  \includegraphics[width=1.0\hsize]{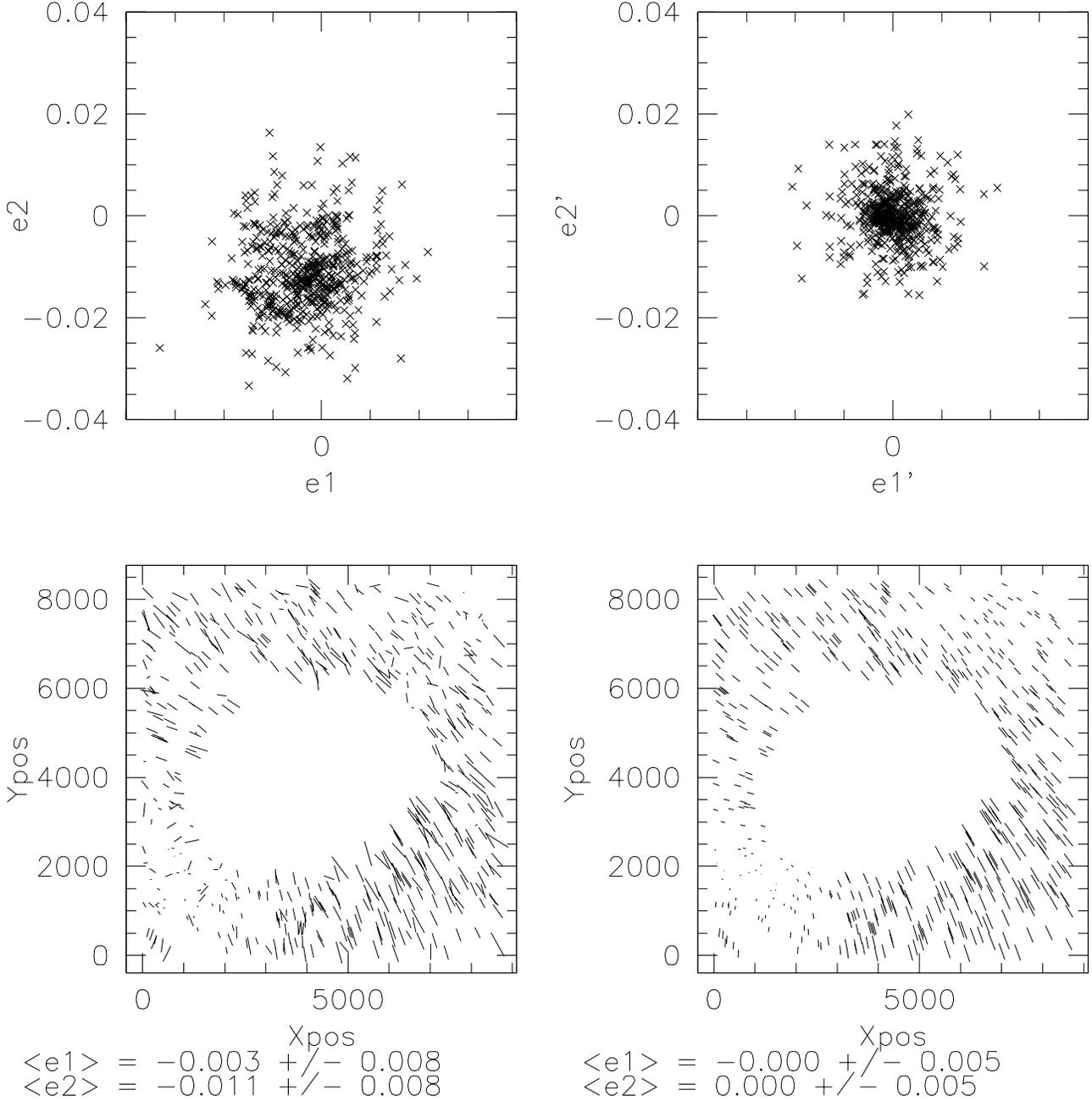}
  \caption{\label{ngc300_aniso}PSF anisotropies in the NGC 300 $R$-band image. Upper left: before 
correction, upper right: after correction. Lower left: Anisotropies as measured in the image,
lower right: a two-dimensional polynomial fit to the PSF anisotropies}
\end{figure*}
\subsection{Data reduction and catalogue creation}
The NGC 300 data was reduced essentially in the same way as described in Sect. 4. Due to the large 
extent of this galaxy a superflat could only be calculated for the field outside NGC 300. Pixels 
lying inside the galaxy were only corrected for the gain differences, which were determined from the 
unaffected outer area. The same held for the sky subtraction of individual images before the coaddition 
process. The sky was modeled outside NGC 300 and assumed to be constant inside, so that no 
discrete jumps appeared between the inner and outer part. This image was then smoothed with a large 
kernel and subtracted. A substantial part of the images suffered from secondary scattering light, 
and from occasional vignetting caused by the filter holder. Most affected was the south-eastern corner 
of the field, where CL0056.02 is located. Our analysis therefore mainly concentrates on CL0056.03.

The photometric zeropoint of the coadded $V$-band image was determined by matching stellar magnitudes 
to the secondary standard stars established by \citet{pietrzyn_standard} in the Johnson-Cousins system. 
No calibration was available for the $R$-band, for which a zeropoint was determined based on the 
expected $V-R$ colour for the red sequence of CL0056.03 ($z=0.165$). We estimate it to be accurate 
within 0\mymag1, which is sufficient for the analysis presented here since it does not rely on 
highly accurate absolute photometry.

For the production of colour catalogues the coadded images were normalised so that their mean
background noise $\sigma_{\rm{back}}=1$. They were then coadded without further rescaling, yielding a 
high signal-to-noise detection image. For this stack an adapted weight map was created combining
the individual weight images accordingly. For the creation of a colour catalogue we used the detection 
image, the detection weight map and the unnormalised images from filters $V$ and $R$, as an input for 
\textit{SExtractor}. Thus the flux for objects in different filters was measured within the same aperture, 
yielding relative colour information for galaxies with good internal accuracy. A further advantage of this 
approach was that the objects detected in the single catalogues for the $V$- and $R$-bands were already in 
the same order and easily merged into one colour catalogue. We calculated the galaxy colours from isophotal 
magnitudes (MAG\_ISO), and used MAG\_AUTO for the magnitudes themselves.
\subsection{Using weak shear for cluster detection}
In the following standard weak lensing notations are used. For a technical review of this topic 
see \citet{bs_review}. The tidal gravitational field of a cluster-sized mass concentration induces 
a coherent distortion pattern in the images of distant background galaxies. By scanning the field 
for such characteristic distortion patterns one can detect mass concentrations directly, irrespective 
of their luminosity \citep[see][for examples]{erben_darkclump,umf00,wtm01,meh02}.

\begin{figure}[t]
  \includegraphics[width=1.0\hsize]{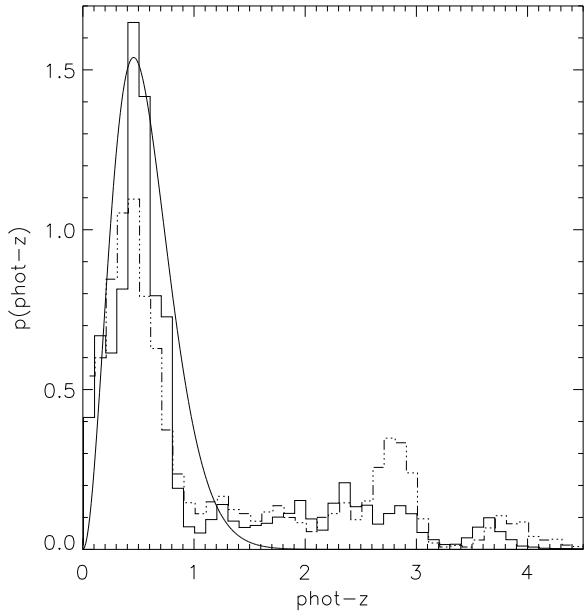}
  \caption{\label{zphot_hist}Photometric redshift distribution for the Chandra Deep Field South,
determined from $UBVRI$ WFI@2.2 photometry. The redshifts were estimated with \textit{hyperz} (solid 
line) \citep{hyperz}. Only objects that had a good redshift fit of $P(\chi^2)>0.9$ went into the 
distribution shown. The dash-dotted line shows an independent measurement of the photometric 
redshifts from E. Hatziminaoglou (ESO) for the same data set (private communication). The peak at 
$z=2.8$ is due to a degeneracy between low and high redshift galaxies and appeared since no prior 
for the luminosity of galaxies was used. These excess objects are randomly drawn from the low-z 
regime. Thus the main peak of this distribution was lowered, but not shifted in redshift. Equation 
(\ref{blainerd_dist}) was fitted to both distributions, yielding consistent values 
$z_{0}=0.373\;(0.377)$.}
\end{figure}

We use the \textit{aperture mass statistics} $M_{\mathrm{ap}}$ \citep{schneider_map} for the 
detection of galaxy clusters. $M_{\mathrm{ap}}$ is a filtered integral of the projected mass 
distribution, $\kappa$, inside an aperture. Its definition reads
\begin{equation} 
M_{\rm ap}=\int_0^{\theta} {\rm d}^2\mytheta \,\kappa(\mytheta, z_{\rm d}, z_{\rm s})\, U(\vartheta)\;.
\end{equation} 
$z_{\rm d}$ and $z_{\rm s}$ are the lens and source redshifts, and $U(\vartheta)$ is a compensated 
filter, i.e. $\int_0^{\theta}d\vartheta\,\vartheta\,U(\vartheta)=0$. The filter function $U(\vartheta)$ 
is chosen as
\begin{equation}
U(\vartheta)=\frac{9}{\pi \theta^2}\left(1-\left(\frac{\vartheta}{\theta}\right)^2\right)
                \left(\frac{1}{3}-\left(\frac{\vartheta}{\theta}\right)^2\right)\;.
\end{equation}
By defining the new filter function 
\begin{equation}
   Q(\vartheta)=\frac 2{\vartheta^2}\int_0^{\vartheta} d\vartheta' \vartheta' U(\vartheta')-U(\vartheta)\;,
\end{equation}
$M_{\rm ap}$ can be expressed in terms of the tangential shear $\gamma_\mathrm{t}$, for which the
observable ellipticities $\epsilon_{\rm t}$ of the background galaxies are an unbiased estimator, as
\begin{equation}
   M_{\rm ap}=\int_0^{\theta} {\rm d}^2\mytheta \gamma_{\rm t}(\mytheta) Q(\vartheta)
             \approx\frac{1}{n} \sum_i\epsilon_{\rm t}(\mytheta_i)Q(\vartheta_i)\;.
\end{equation}
Hence, we can calculate the scalar $M_{\rm ap}$ directly from observables. The noise for 
$M_{\mathrm{ap}}$ is evaluated as
\begin{equation}
\sigma^2_{\rm Map}=\frac{\pi\sigma_{\epsilon}^2}n \int_0^{\theta}{\rm d}\vartheta\,\vartheta\,
                   Q^2(\vartheta) \rightarrow \frac{\sigma_{\epsilon}^2}{2n^2}
                   \sum_i Q^2(\vartheta_i),
\end{equation}
where $\sigma_{\epsilon}$ is the ellipticity dispersion and $n$ the number density of background 
galaxies. We determined $n=20 \;\rm{arcmin}^{-2}$ and $\sigma_\epsilon=0.34$ for the WFI@2.2 data at 
hand. Based on this expression, the S/N ratio for mass peaks can be estimated. For 
doing so, a projected mass distribution $\kappa(\mytheta, z_{\rm d}, z_{\rm s})$ at a redshift 
$z_{\rm d}$ is assumed, together with a redshift distribution $p(z_{\rm s})$ for the lensed 
galaxies. The expected signal $S$ then reads
\begin{equation}
   S=\int_0^{\theta}{\rm d}^2\mytheta\int_{z_{\rm d}}^{\infty} {\rm d}z_{\rm s}\,
   \kappa(\mytheta, z_{\rm d}, z_{\rm s})U(\vartheta)p(z_{\rm s}).
\end{equation}
For the mass distribution a NFW profile \citep{navarro} was used. \citet{kruse}, showed how to 
convert this profile into the projected surface mass density, $\kappa$. For the source redshifts 
we took the normalised distribution 
\begin{equation}
   \label{blainerd_dist}
   p(z_{\rm s})=\frac 3{2z_0}\left(\frac {z_{\rm s}}{z_0}\right)^2
   \exp\left[ -\left(\frac {z_{\rm s}}{z_0}\right)^{1.5}\right]\,,
\end{equation}
proposed by \citet{brainerd}. We fixed the parameter $z_0$ based on photometric redshifts, computed 
from the $UBVRI$ WFI@2.2 data (EIS Deep Public Survey) for the Chandra Deep Field South. The exposure 
time in the $R$ filter was 9000 seconds, lower than the $\sim15000$ seconds for the NGC 300 field. Fig. 
\ref{zphot_hist} shows the photometric redshifts and the fit to the $p(z_{\rm s})$ profile. A value 
of $z_0=0.37$ is determined from the fit. We conservatively increased this value to $z_0=0.4$ for the
S/N computation, given the significantly larger exposure time in $R$.

\begin{figure}[t]
  \includegraphics[width=1.0\hsize]{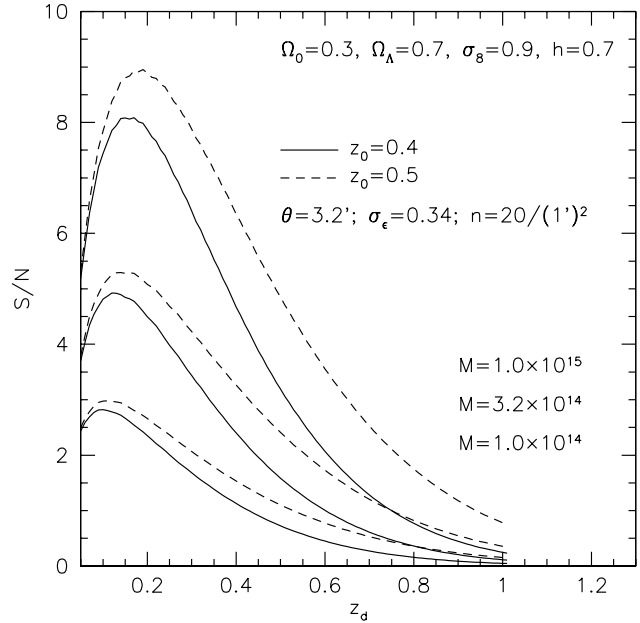}
  \caption{\label{sn_graph}The expected S/N ratios for various massive dark matter haloes, as 
detected with $M_{\rm ap}$ in the NGC 300 WFI@2.2 field. The effect of a background galaxy population 
with higher redshift is shown, too.}
\end{figure}

Fig. \ref{sn_graph} shows some predicted S/N ratios for massive haloes as a function of redshift. From this 
plot we see that clusters with masses of $3.2\times 10^{14} M_\odot$ are detectable up to a redshift 
of $z=0.35\;(0.42)$, and those with $1.0\times 10^{15} M_\odot$ can be found up to $z=0.5\;(0.65)$ for a 
filter size of 3\myarcmin2 (4\myarcmin0, not shown). We are not sensitive to structures with 
$\sim1.0\times 10^{14} M_\odot$ at the 3\myarcmin2 scale, but can detect them at the 4\myarcmin0 scale 
up to $z=0.14$. For comparison, S/N ratios are also shown for the case when background galaxies are at 
somewhat higher redshift ($z_0=0.5$).

The S/N ratio for $M_{\rm ap}$ peaks in a given data field can also be determined from the data 
itself, by randomizing ellipticities while keeping galaxy positions fixed. In the following we make
use of this fact. For the weak lensing analysis the $R$-band image was used. To avoid any biasing of 
the detection algorithm, NGC 300 was masked and replaced by the mean sky background, removing most of 
the flux present in the image. With \textit{SExtractor} all objects with at least 6 connected pixels 
$\ge2\sigma$ above the sky background noise were detected. This catalogue contained 44146 objects. The 
shear estimates for these objects were determined with the KSB algorithm as proposed by \citet{ksb}. 
An extensive description of our approach using the KSB method, including the PSF corrections, can be 
found in \citet{howaccurate}. Figure \ref{ngc300_aniso} shows the measured and the corrected PSF 
anisotropies for this particular field. After filtering, the background galaxy catalogue contained 
12694 objects with securely determined shapes, a detection significance $\nu>12$ (calculated by KSB) 
and a bright magnitude limit of $R>23.0$. This translates to the above mentioned number density of 
$\sim20$ galaxies per square arcmin.

\begin{figure*}[t]
  \includegraphics[width=1.0\hsize]{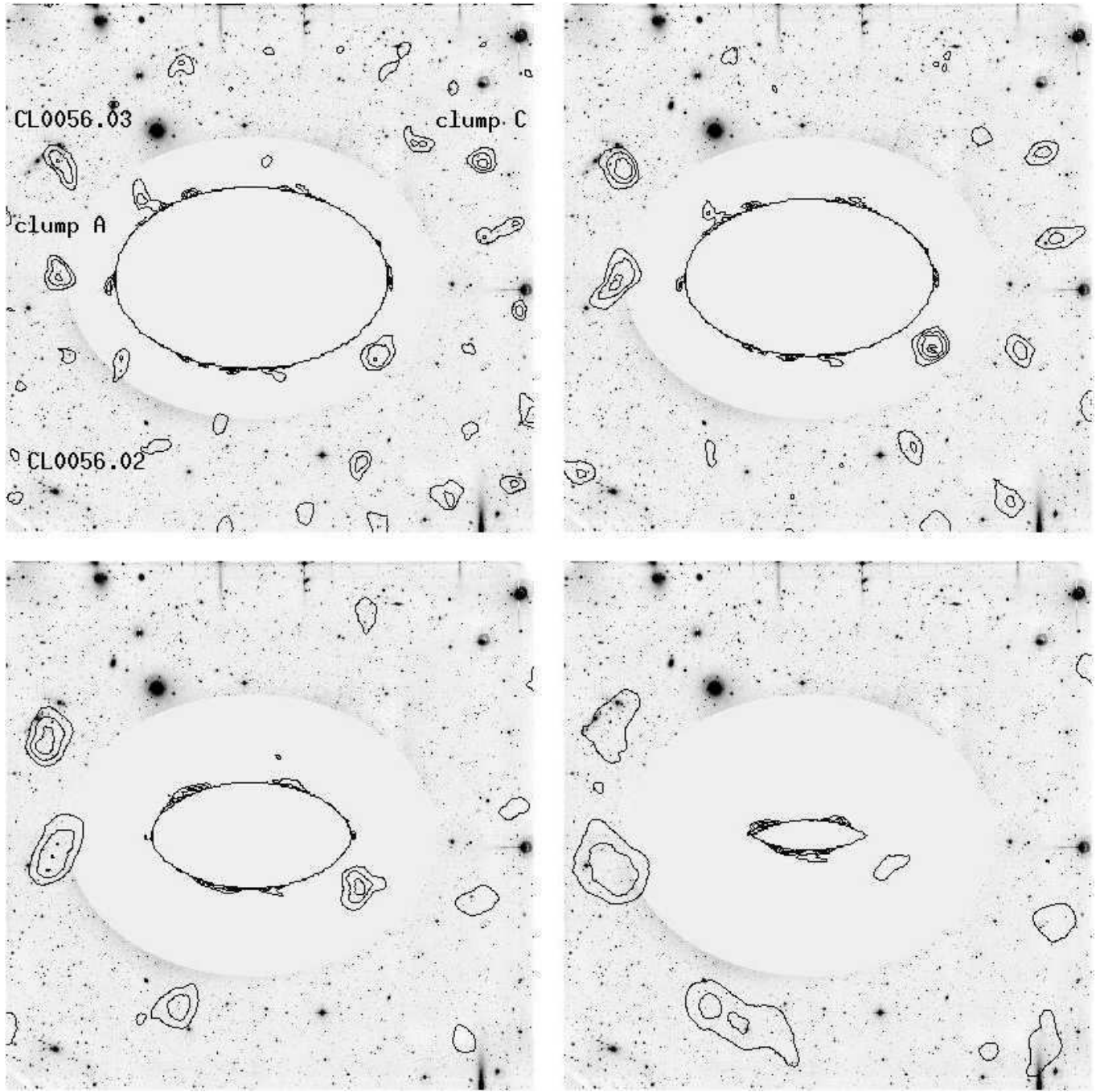}
  \caption{\label{peaks}The $M_{\rm ap}$ statistics for various filter scales: 3\myarcmin2 (upper left), 
4\myarcmin0 (upper right), 5\myarcmin6 (lower left) and 8\myarcmin0 (lower right). Shown are $M_{\rm ap}$ 
probability contours drawn from 5000 randomisations each. The contours depict the 2.0, 2.5, 3.0, 3.5$\sigma$ 
levels. The cluster CL0056.03 can be found left of the brightest star, and CL0056.02 sits in the lower 
left corner. See also Figs. \ref{redsequence_total} and \ref{redsequence_redder} for comparison.
The large elliptical contour arises from the fact that $M_{\rm ap}$ can not determine any value for the 
pixels inside the contour, since for those no galaxies lie inside the filter. Thus, the distance from this 
ellipse to the outer edge of the mask depicts the radius of the $M_{\rm ap}$ filter function.}
\end{figure*}

Fig. \ref{peaks} shows the $M_{\rm ap}$ results for various filter widths. As can be seen, we recovered
CL0056.03 in all but the largest filter scale at a 3$\sigma$ level within 50--100\myarcsecnodot 
south of the cluster centre, whereas CL0056.02 was not detected. For the latter cluster useful shear 
information could only be obtained from about 30\% of the area which would be available if the cluster 
did not lie next to the field corner in a region with bad image quality. This increased the 
noise in $M_{\rm ap}$ by a factor of $\sim3$. The argument of field truncation also holds for CL0056.03, 
but to a much lesser extent. There the useable field was limited by the edge of the image, 2\myarcmin5 
east of the cluster centre, and NGC 300 5\myarcmin5 south-west. From our previous S/N considerations we
conclude that a mass of $\sim1.4\times 10^{14} M_{\odot}$ at redshift of $z=0.165$ would produce a
comparably significant $M_{\rm ap}$ detection.

The $M_{\rm ap}$ statistics furthermore picks up a number of other peaks. In the upper left panel of 
Fig. \ref{peaks}, the one for the smallest filter scale, we find a $3\sigma$ peak that drops to 
$2.5\sigma$ in the upper right panel, and vanishes for larger filter scales (clump `C'). Another peak 
below CL0056.03 is detected at the $3\sigma$ level for all but the largest filter scale (clump `A'). 
Both peaks lie within 50\myarcsecnodot and 20--70\myarcsecnodot of two clumps of red galaxies, as is 
shown in the next section. There we determine their redshifts $z_{\rm{d,A}}=0.47$ and $z_{\rm{d,C}}=0.43$. 
Our S/N calculations then yield masses of $M_{\rm{A}}(4\myarcmin0)\approx(4\pm3)\times10^{14}M_{\odot}$ and
$M_{\rm{C}}(3\myarcmin2)\approx(6\pm3)\times10^{14}M_{\odot}$ for these two clumps and for the filter scales
in which they are detected most significantly. See also Fig. \ref{redsequence_redder} for those two 
concentrations of red galaxies.

Besides, a stable peak is found inside the mask of NGC 300, with a detection limit of 
$3-3.5\sigma$ for the three smaller filter scales. Such a detection is not surprising, since 
$M_{\rm ap}$ is a highly non-local measure. It can pick up those parts of the shear field 
from a possible cluster hidden behind NGC 300 that extend beyond NGC 300 itself. Shear fields 
for massive clusters of galaxies have been traced beyond 10\myarcminnodot of cluster centres \citep[see]
[for examples]{clowe}, thus a sufficiently massive cluster behind NGC 300 could easily be 
picked up. The detection in question is at the very outer edge of the galaxy disk, but the 
confusion limit of foreground stars in NGC 300 is already reached. Yet the optical thickness 
of the disk is still small, so that three larger and brighter isolated galaxies can be seen 
within a 2\myarcminnodot wide window. Thus, if there was a massive lower redshift cluster 
such as CL0056.03 at this position, it could be seen through the disk.
\begin{figure}[t]
  \includegraphics[width=1.0\hsize]{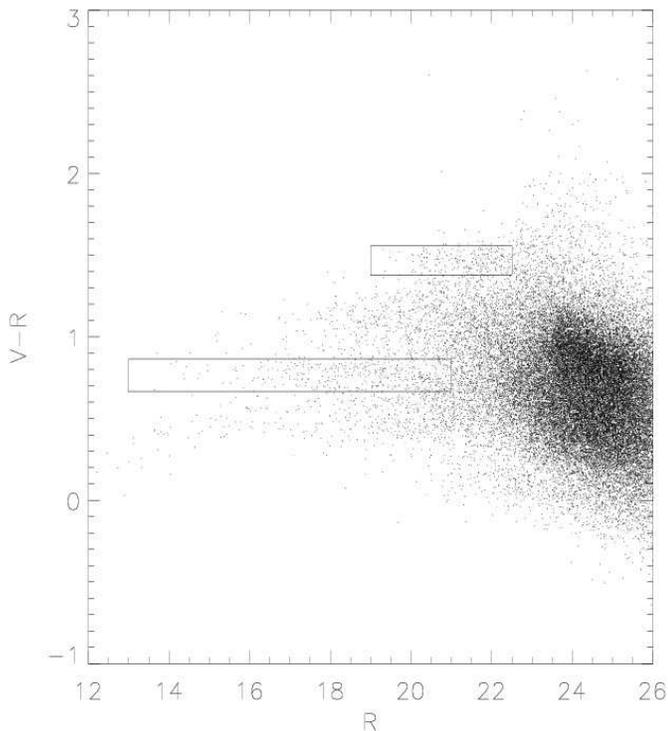}
  \caption{\label{redsequence_colmag}The $(V-R,R)$-colour-magnitude diagram for all galaxies around 
NGC 300. The selected cluster sequence for CL0056.03 and CL0056.02 is marked by the large box.
We do see clustering of redder objects in the upper window also. See the text for more details.}
\end{figure}
There is no indication for such an object at this position. More distant clusters, however, could no 
longer be identified as such, since their smaller and fainter images are drowned in the foreground 
confusion. No excess X-Ray emission is found in the ROSAT All Sky Survey \citep{rosat} for the 
$M_{\rm ap}$ detections, apart from CL0056.03.
\subsection{Using the red sequence for cluster detection}
\begin{figure*}[t]
  \includegraphics[width=1.0\hsize]{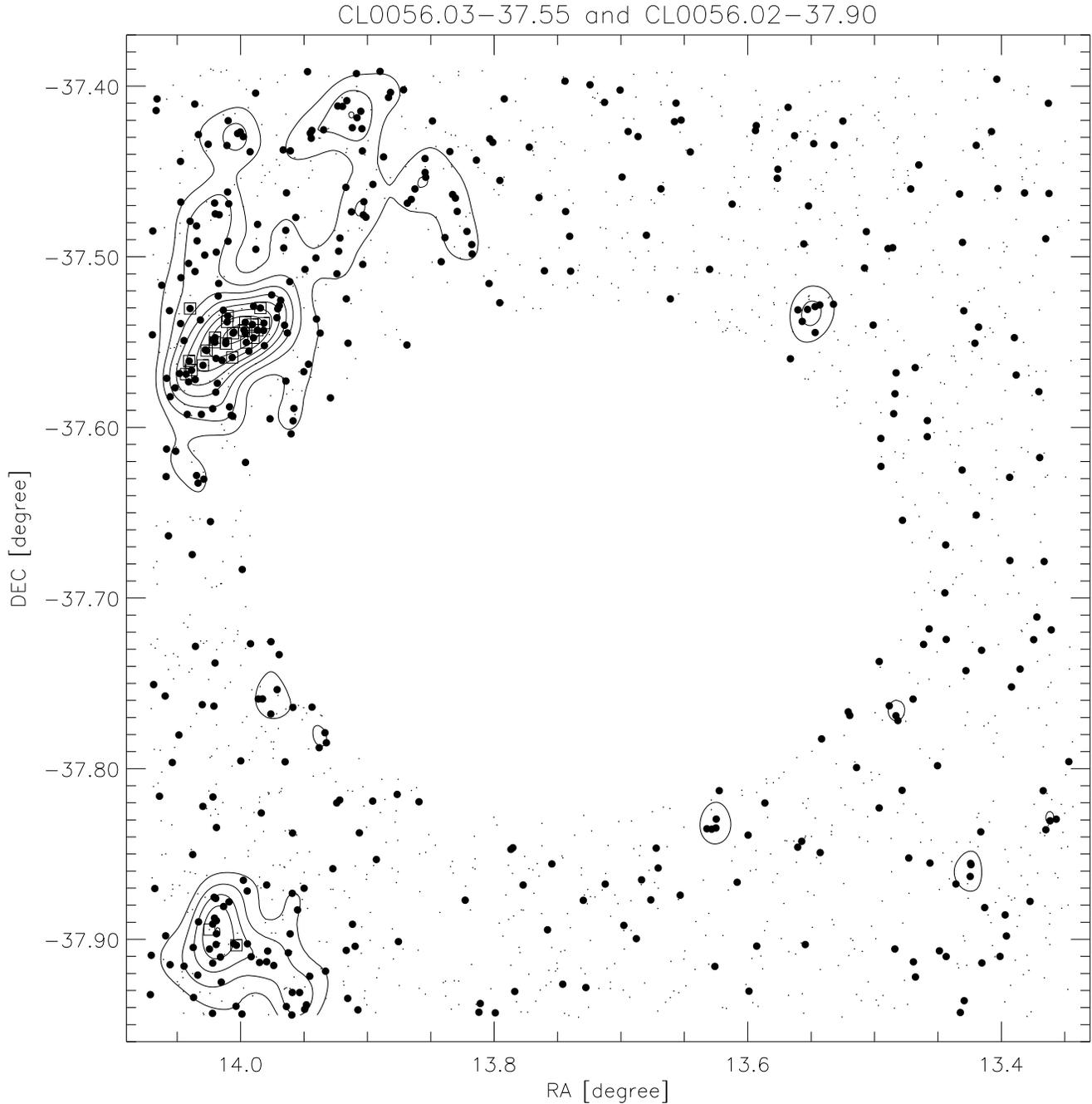}
  \caption{\label{redsequence_total}Galaxies inside the lower red cluster sequence of 
Fig. \ref{redsequence_colmag} are shown as big dots. The small dots indicate galaxies brighter 
than $R=21$ which do not fall inside the red sequence window. The overlaid contours are isodensity 
contours for the red sequence members, smoothed at a 3\myarcmin6 scale, and starting with the 
$1\sigma$-overdensity contour in steps of $1\sigma$. CL0056.03 and CL0056.02 are detected at the 
$8\sigma$ and $5\sigma$ level respectively. CL0056.03 appears strongly elongated, with an intersecting 
12\myarcminnodot long filament extending north-south at its eastern side. Part of the filament could 
belong to Abell S0102 at $(\alpha,\delta)=(13.91,-37.41)$, a poor cluster at $z=0.05$. Galaxies with 
measured spectra are highlighted with small squares around them.}
\end{figure*}

In a $(V-R,R)$ colour-magnitude diagram (Fig. \ref{redsequence_colmag}), plotted for all galaxies 
in the vicinity of NGC 300, we loosely selected a red cluster sequence with $0.67<V-R<0.87, 
13.0<R<21.0$. The sky distribution for objects inside this window is plotted in Fig. \ref{redsequence_total} 
(big red dots), whereas all other objects with $R<21$ and outside the window are shown as small 
dots. We then calculated the rms of the projected density of all red sequence members, and overlaid
isodensity sigma-contours for their distribution. The smoothing length for the density was 3\myarcmin6.
We recover CL0056.03 and CL0056.02 at the $8\sigma$ and $5\sigma$ level in overdensity, respectively.
Thus the cluster sequence we selected in the colour-magnitude diagram is actually a merger of two 
clusters of galaxies.

\begin{figure}[t]
  \includegraphics[width=1.0\hsize]{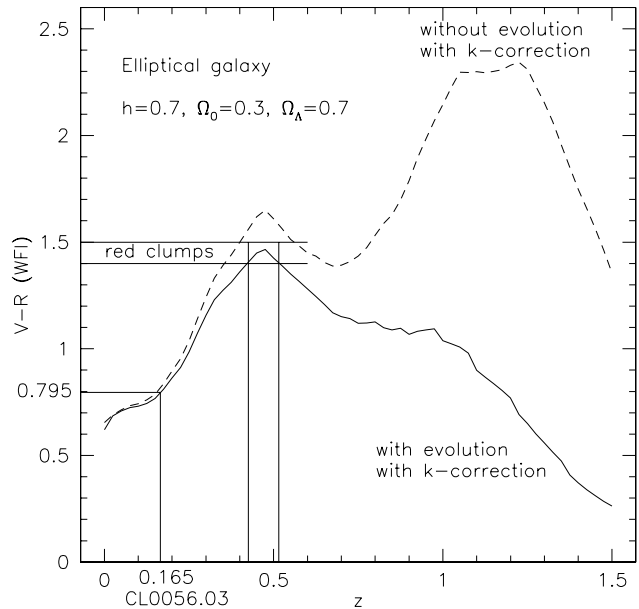}
  \caption{\label{ellipticaltracks}Shown are predictions for the WFI@2.2 $V-R$ colours of elliptical
galaxies as a function of redshift \citep{hyperz,bruzual}. Based on the track that includes 
evolutionary effects we estimate the redshifts for the shear-selected red clumps.}
\end{figure}
The remaining galaxies that lied within the red sequence window, but which did not belong
to either of the two clusters, did not clump throughout the rest of the field on a level higher than 
$1\sigma$. Furthermore, all galaxies brighter than $R=20$ and outside the red sequence window 
showed no clumping at a level higher than $2\sigma$. We additionally checked the clustering properties 
of all objects brighter than $R=23$, and did not find any highly significant clumping apart from 
galaxies with $1.37<V-R<1.56, 19.0<R<22.5$. There we found four significant overdensities `A' to `D', 
with `A' and `C' coincident with $M_{\rm ap}$ detections (see above). Table \ref{clump_results}
summarises the properties and redshift estimates based on the track for elliptical galaxies in
Fig. \ref{ellipticaltracks}.
\begin{figure*}[t]
  \includegraphics[width=1.0\hsize]{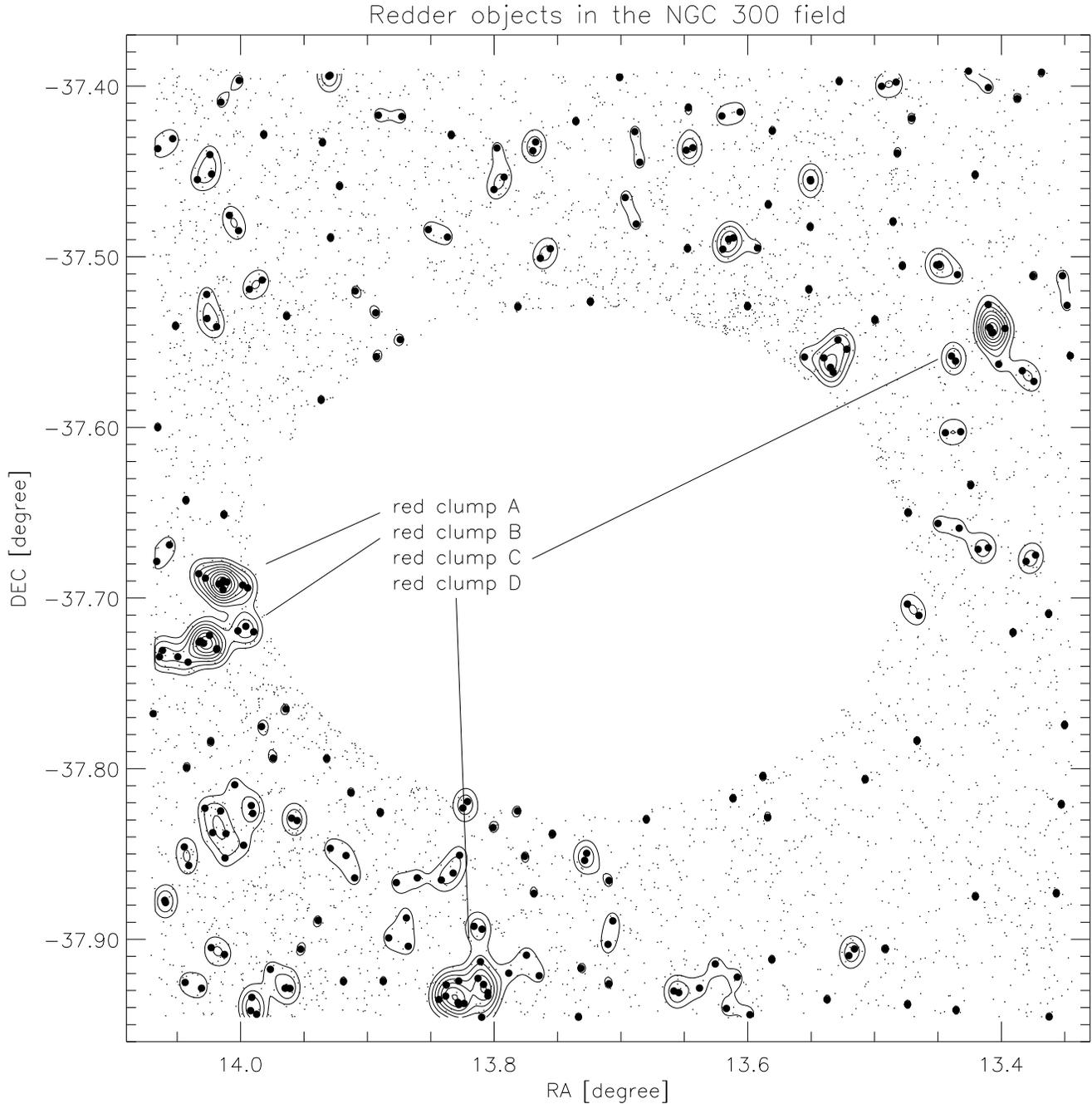}
  \caption{\label{redsequence_redder}Shown is the distribution of galaxies inside the upper red 
cluster sequence of Fig. \ref{redsequence_colmag} (big dots). The small dots indicate galaxies 
brighter than $R=22.5$ which do not fall inside the red sequence window. Clump `A' is a very tight 
concentration of nine galaxies with $R<22.3$ within 25\myarcsecnodot. There are five more galaxies 
spread along a 1\myarcminnodot filament to the east and to the west. Clump `B' is a much looser, 
4\myarcminnodot long filament consisting of 13 galaxies without a central concentration. In the 
centre of the filament one finds a $R=19.4$ bright elliptical, about 1 magnitude brighter than the 
second brightest member of this concentration. Clump `C' consists of 6 galaxies within 80\myarcsecnodot, 
the brightest one with a R-magnitude of 20.9. Finally, clump `D' is a very loose grouping of fainter 
red galaxies that looks like a chance alignment rather than a cluster. Note that the resolution of this 
plot is not high enough to reveal all candidate galaxies. The smoothing length for the density contours 
was 1\myarcmin5 arcmin. Clumps `A' and `C' lie within $\sim50\myarcsecnodot$ of weak lensing peaks.}
\end{figure*}
\begin{table}[ht]
   \caption{\label{clump_results}All clumps except `C' lie close to the peak of the track 
   in Fig. \ref{ellipticaltracks}, reducing the ambiguity in their redshift estimate.}
   \begin{tabular}{l|lll}
   \hline 
   \noalign{\smallskip}
   Red clump  & $\langle V-R \rangle$  & $z_{\rm{est}}$\\
   \noalign{\smallskip}
   \hline 
   A & $1.46\pm0.02$ & $0.47\pm0.05$\\
   B & $1.50\pm0.03$ & $0.47\pm0.05$\\
   C & $1.41\pm0.02$ & $0.43\pm0.05, 0.51\pm0.05$\\
   D & $1.50\pm0.03$ & $0.47\pm0.05$\\  
   \noalign{\smallskip}
   \hline 
   \noalign{\smallskip}
   \noalign{\smallskip}
   \end{tabular}
   \small
\end{table}
We see that all clumps lie at similar redshifts, with some ambiguity for clump `C'. Actually, peaks 
`A' and `B' could form a connected system if spectroscopically confirmed to be at the same redshift. 
More details are given in the caption of Fig. \ref{redsequence_redder}.

\begin{figure*}[t]
  \includegraphics[width=1.0\hsize]{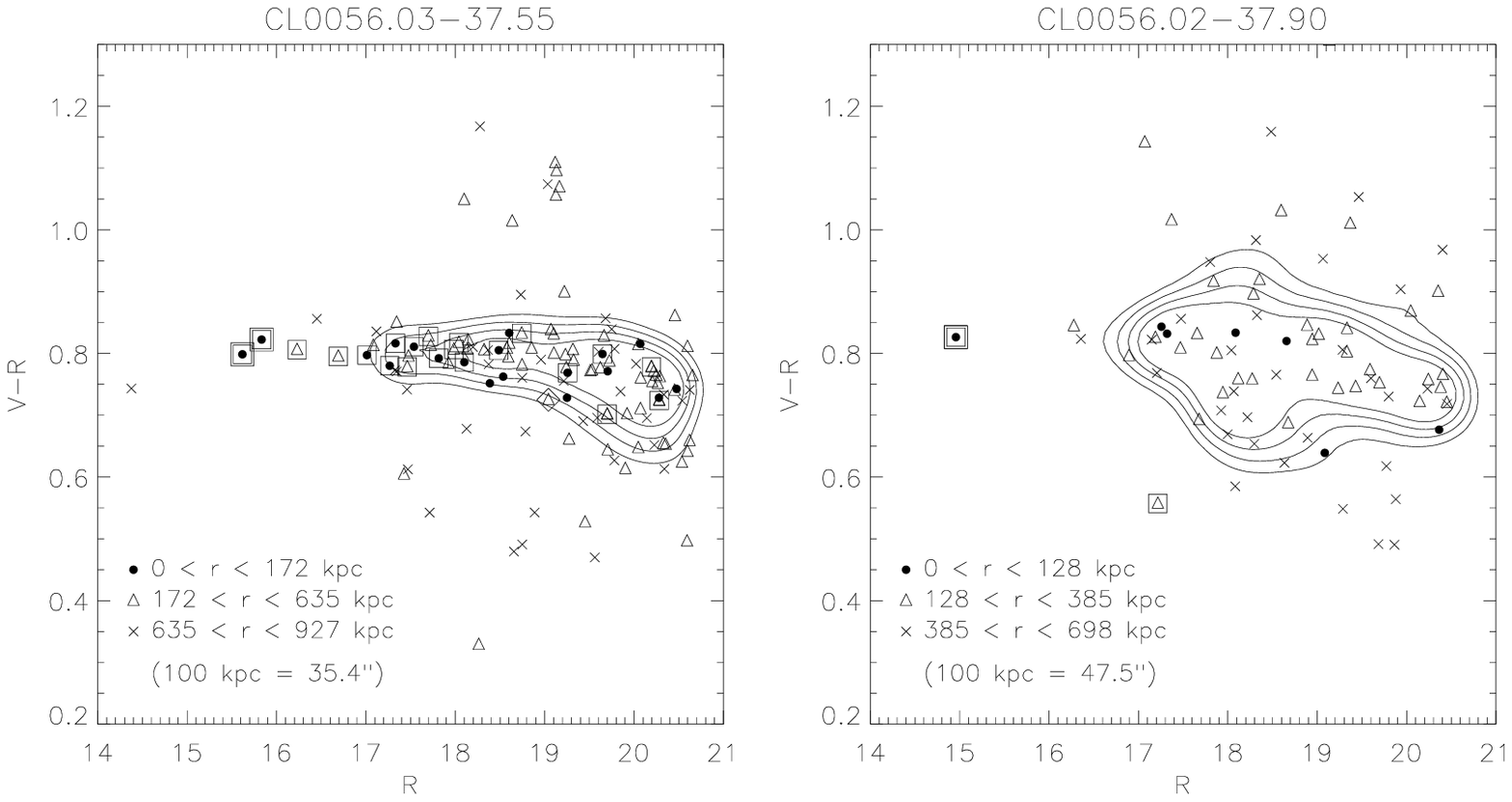}
  \caption{\label{cl0053_colmag}Colour-magnitude diagrams for galaxies inside the cluster sequence 
window. Galaxies with spectroscopically determined redshifts are marked with squares, the cD galaxies 
are indicated with a double square (they also have measured redshifts). The galaxy with $R=19$ and 
$V-R=0.73$, marked with a diamond symbol, has $z=0.27$, higher than the cluster redshift. Furthermore, 
galaxies were split into three distance bins as seen from the geometrical cluster centre. The outer radii 
(635 kpc respectively 385 kpc) of the second annuli were chosen in a way that galaxies in the third, 
outer annulus do not show an apparent concentration with respect to the red cluster sequence any more. 
Isodensity contours in the colour-magnitude space were calculated from all galaxies.}
\end{figure*}
\subsection{Butcher-Oemler effect in CL0056.03}
Fig. \ref{cl0053_colmag} depicts the two colour-magnitude diagrams for CL0056.03 and CL0056.02. 
Shown are galaxies with $V-R=0.80\pm0.5$, $13.0<R<21.0$ and within 5\myarcmin5 of the cluster 
centres. These galaxies were split into three distance bins (see Fig. \ref{cl0053_colmag}), 
in order to detect a possible evolution in their colour as a function of distance from the 
cluster core. The number of interlopers is small; we counted 41 galaxies with the same properties 
in a $6\myarcmin 5\times 33\myarcminnodot$ strip on the opposite side of NGC 300, corresponding to 
a number density of about 0.19 galaxies per square arcmin. If no further clumping takes place 
along the line-of-sight, 11\% (19\%) of interlopers can be expected within the area under 
consideration for CL0056.03 (CL0056.02).

For CL0056.03, in the innermost bin 11 out of 19 galaxies have measured redshifts, in the second 
bin it is 10 out of 60 galaxies, and in the outermost bin none of the 34 galaxies have known 
redshifts. 10 galaxies appear to lie above the upper envelope ($V-R>0.87$) of the cluster sequence, 
7 of those are from the second distance bin, and the rest comes from the third bin. Their number 
is in agreement with the expected amount of interlopers. 4 of those 10 galaxies are spatially 
concentrated within 1\myarcmin2. None of the 10 has spectroscopically determined redshifts, 
we assume that they are in the background of CL0056.03 since they lie beyond the upper envelope 
of $V-R<0.87$ as defined by the red sequence.

A closer inspection of the colour-magnitude diagram for CL0056.03 reveals several features:
\begin{itemize}
\item{Galaxies within 172 kpc of the cluster core (red dots) follow a tight correlation of
\begin{equation}
V-R=(-0.013\pm0.004)R+1.028\pm0.077
\end{equation}}
\item{Galaxies in the range [172--635] kpc of the cluster core (green triangles) show a steeper 
correlation with substantially larger scatter:
\begin{equation}
V-R=(-0.030\pm0.014)R+1.376\pm0.264\,.
\end{equation}
Less luminous galaxies with $R>19$ are disproportionately bluer than those with $R<19$, with the latter 
ones lying on the same sequence as the galaxies inside the innermost annulus. Fainter galaxies inside 
172 kpc are bluer too, but to a much smaller extent than the galaxies in the second distance bin.}
\item{Galaxies within 635 kpc and $R<19$ have $V-R=0.795\pm0.02$, showing no evolution.}
\item{Galaxies in the range [635--927] kpc show no obvious concentration in colour-magnitude space, 
but the red sequence appears as an upper limit for the colour of those galaxies. Only two of them 
show significantly redder colours, whereas about a dozen appear bluer and about 20 lie on the red sequence 
as defined by galaxies in the inner two annuli. The upper limit indicates that a significant fraction of 
galaxies further away than 635 kpc could belong to the cluster population. However, without any measured 
redshifts this fraction can not be quantified at this stage.}
\end{itemize}
\begin{table}[t]
   \caption{\label{colmag_conclusions}The Butcher-Oemler Effect. In the case of CL0056.03 a galaxy is 
defined to be `blue' if $V-R<0.755$, i.e., if it lies more than $2\sigma$ below the cluster sequence 
$V-R=0.795\pm0.02$, as defined by galaxies with $R<19$. Galaxies with $V-R>0.87$ are exluded from the 
statistics. For CL0056.02 the $2\sigma$-threshold for `blue' galaxies is $V-R<0.79$. Galaxies with 
$V-R>0.85$ are excluded from the statistics. Note that the number of interlopers is reduced in the 
outer two annuli since they are significantly truncated by the edge of the field of view.}
   \begin{tabular}{l|lll}
   \hline 
   Distance from  & Galaxies    & Fraction of   & est. No. of\\
   cluster centre & with(out) z & blue galaxies & interlopers\\
   \hline 
   \noalign{\smallskip}
   \noalign{\smallskip}
   CL0056.03 & & &\\
   \hline 
   0--200 kpc    & 11 (8)  & 0.26 & 0.6\\
   200--740 kpc  & 10 (50) & 0.35 & 6.8\\
   740--1080 kpc & 0 (34)  & 0.57 & 6.5\\
   \hline 
   \noalign{\smallskip}
   \noalign{\smallskip}
   CL0056.02 & & &\\
   \hline 
   0--143 kpc    & 1 (6)   & 0.29 & 0.6\\
   143--430 kpc  & 1 (26)  & 0.59 & 4.5\\
   430--780 kpc  & 0 (23)  & 0.83 & 6.2\\
   \hline 
   \noalign{\smallskip}
   \noalign{\smallskip}
   \end{tabular}
   \small
\end{table}
To summarise, for CL0056.03 a clear colour evolution for $R>19$ and inside 635 kpc is found, increasing 
with the distance from the cluster core. Furthermore there exists a population of galaxies beyond 635 
kpc which appears to share the same evolution limit as galaxies further inside, but they do not form a 
clear cluster sequence. The fraction of blue galaxies increases towards larger radii.

For CL0056.02 we find similar, but less significant results. The cluster sequence is broader
and not so well-defined than the one for CL0056.03. There is no clear cut-off towards redder colours. 
Only galaxies inside the innermost distance bin and with $R<19$ form a tight sequence with 
$V-R=0.831\pm0.009$. Galaxies in the second bin have a considerably larger scatter for brighter 
magnitudes, but a smaller one at the faint end as compared to CL0056.03. Objects within 385 kpc show
a clear colour evolution with a slope of $-0.029\pm0.019$ for $R>18.5$. Beyond 385 kpc the galaxy 
population becomes significantly bluer, 83\% of them lie below the red sequence. The fraction of blue 
galaxies is larger than for CL0056.03 in the outer two annuli, whereas in the cluster cores 
their fraction is comparable. We note that the data quality in the field around CL0056.02 is 
significantly lower than for CL0056.03, increasing the photometric errors. Table \ref{colmag_conclusions} 
summarises the Butcher-Oemler effect found for both clusters.
\subsection{Substructure in CL0056.03}
\begin{figure*}[t]
  \includegraphics[width=1.0\hsize]{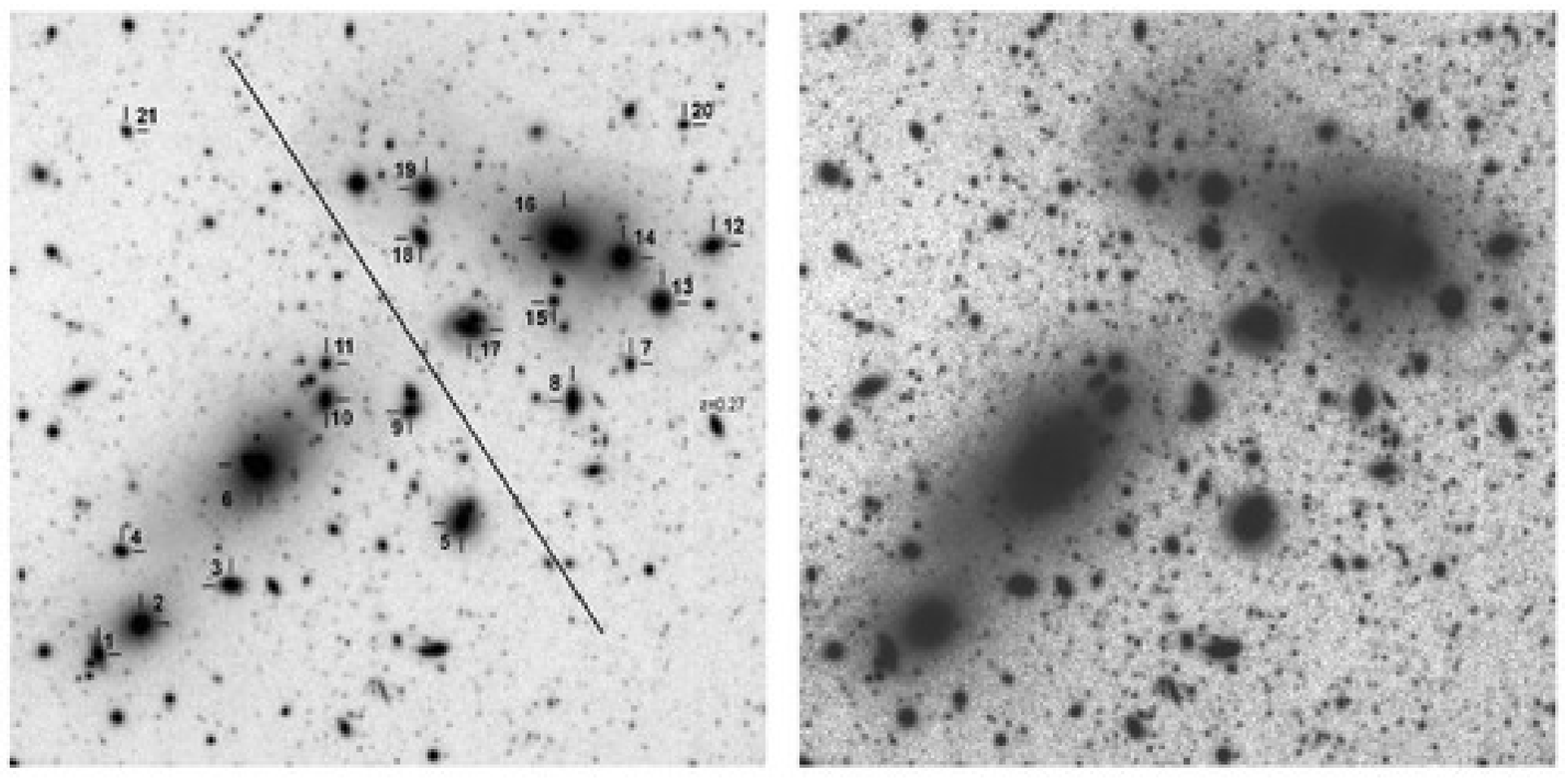}
  \caption{\label{cl0053_clusterimages}Image of CL0056.03. In the left panel the 21 
spectroscopically confirmed cluster members are shown, together with the perpendicular 
bisector of the connection line between the two cD galaxies, which we chose to split 
the cluster galaxies into two samples. An [OII] emission line galaxy at higher redshift 
is indicated. The right panel shows a steeper scaled version of the image at left, showing
that $\mathrm{cD_1}$ and $\mathrm{cD_2}$ are both embedded in large, but well separated haloes.}
\end{figure*}

In CHM98 a velocity dispersion of $1144^{+234}_{-145} \mathrm{km\,s^{-1}}$ is stated for CL0056.03, 
based on the redshifts in Table \ref{redshifts}. From this dispersion the authors calculated a 
mass of $\sim 2\times10^{15}\,\mathrm{M_\odot}$, assuming a virialised cluster and spherical symmetry. 
This value places CL0056.03 in the same league as supermassive clusters such as Abell 1689. Looking at 
Fig. \ref{redsequence_total} one notices the large elongation ($\sim1:4$) in the distribution of the 
red sequence members, arguing against spherical symmetry and relaxation, in which case the virial 
theorem yields inaccurate results. Given the large angular separation of $105''$ (300 kpc) of the 
two cD galaxies, we searched for a correlation between the positions of galaxies and their redshifts, 
indicating an ongoing merging process. The galaxies were split into two samples by drawing the 
perpendicular bisector of the connection line between the two cD galaxies (see left panel in Fig. 
\ref{cl0053_clusterimages}). This way 10 galaxies with redshifts were assigned to the eastern 
cD galaxy (hereafter $\mathrm{cD_1}$), and 11 to the western one ($\mathrm{cD_2}$).
\begin{table}[t]
   \caption{\label{redshifts}Heliocentric redshifts for CL0056.03, taken from CHM98. Here,
   $v$ denotes the radial velocity in $\mathrm{km\,s^{-1}}$ together with its measurement errors
   $\sigma_v$, and the index $m$ indicates whether a galaxy is closer to the eastern cD galaxy 
   ($\mathrm{ID}=6$), or to the western one ($\mathrm{ID}=16$).} 
   \scriptsize
   \begin{tabular}{lllllll}
   \hline 
   \noalign{\smallskip}
    ID & RA (2000.0)& DEC (2000.0)& v & $\sigma_v$ & z & m\\
   \noalign{\smallskip}
   \hline 
   \noalign{\smallskip}
    1  &  00:56:10.23  &  --37:34:07.9  &  51570  &  282  &  0.17202  &  1 \\
    2  &  00:56:09.22  &  --37:33:59.5  &  50799  &   51  &  0.16945  &  1\\
    3  &  00:56:07.07  &  --37:33:49.4  &  51628  &  138  &  0.17221  &  1 \\
    4  &  00:56:09.71  &  --37:33:39.9  &  50709  &  120  &  0.16915  &  1 \\
    5  &  00:56:01.49  &  --37:33:31.6  &  49315  &  111  &  0.16450  &  1 \\
    6  &  00:56:06.41  &  --37:33:18.0  &  49754  &  154  &  0.16596  &  cD 1 \\
    7  &  00:55:57.50  &  --37:32:51.5  &  46839  &   38  &  0.15624  &  2 \\
    8  &  00:55:58.90  &  --37:33:00.7  &  48490  &   50  &  0.16175  &  2 \\
    9  &  00:56:02.76  &  --37:33:03.9  &  49588  &  192  &  0.16541  &  1 \\
   10  &  00:56:04.76  &  --37:33:00.3  &  50394  &  117  &  0.16810  &  1 \\
   11  &  00:56:04.75  &  --37:32:51.0  &  51594  &   84  &  0.17210  &  1 \\
   12  &  00:55:55.59  &  --37:32:20.2  &  48514  &  145  &  0.16183  &  2 \\
   13  &  00:55:56.74  &  --37:32:35.4  &  49555  &  123  &  0.16530  &  2 \\
   14  &  00:55:57.74  &  --37:32:23.4  &  48256  &  101  &  0.16096  &  2 \\
   15  &  00:55:59.32  &  --37:32:35.2  &  49041  &  145  &  0.16358  &  2 \\
   16  &  00:55:59.09  &  --37:32:18.2  &  48888  &  117  &  0.16307  &  cD 2 \\
   17  &  00:56:01.32  &  --37:32:41.9  &  50168  &  187  &  0.16734  &  2 \\
   18  &  00:56:02.45  &  --37:32:18.0  &  47381  &  138  &  0.15805  &  2 \\
   19  &  00:56:02.41  &  --37:32:05.7  &  49097  &  113  &  0.16377  &  2 \\
   20  &  00:55:56.18  &  --37:31:48.3  &  49439  &   14  &  0.16491  &  2 \\
   21  &  00:56:09.54  &  --37:31:49.0  &  49863  &  107  &  0.16633  &  1 \\
   \noalign{\smallskip}
   \hline 
   \noalign{\smallskip}
   \noalign{\smallskip}
   \end{tabular}
   \small
\end{table}
\begin{figure}[t]
  \includegraphics[width=1.0\hsize]{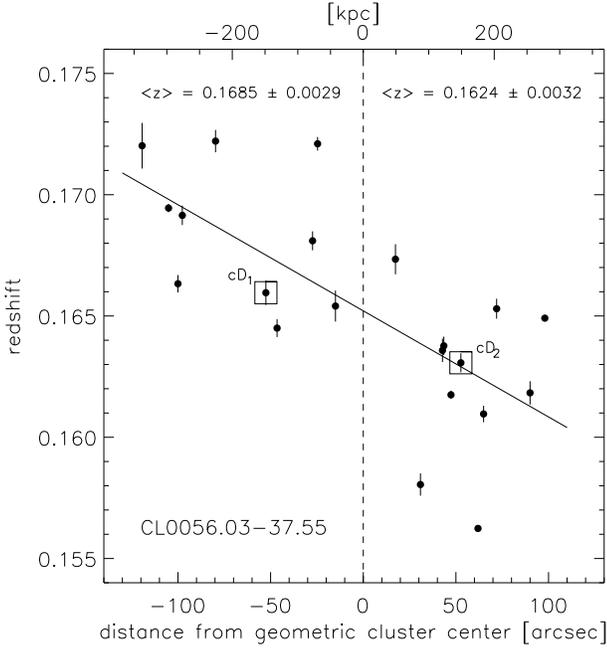}
  \caption{\label{cl0053_z-spacecorr}The redshifts of the galaxies in CL0056.03 as a function of 
distance from the geometrical cluster centre. There is a highly significant ($\sim 4\sigma$) 
correlation between galaxy position and redshift, indicating that CL0056.03 actually consists of 
two possibly merging subclumps. The slope of the linear fit is different from zero on the 
$4.1\sigma$ level. Note that $\mathrm{cD_1}$ has a peculiar motion of 780 $\rm{km\,s^{-1}}$ with 
respect to its own clump.}
\end{figure}
Fig. \ref{cl0053_z-spacecorr} shows the redshifts of galaxies as a function of distance from the 
geometric cluster centre, defined as the centre of the connection line between $\mathrm{cD_1}$ and 
$\mathrm{cD_2}$. We find different mean redshifts for the galaxies around $\mathrm{cD_1}$ 
and $\mathrm{cD_2}$, translating into a velocity difference of $\sim1824\,\rm{km\;s^{-1}}$.
The velocity dispersions of the two clumps are thus significantly smaller than the one given in
CHM98 for the whole system, as is the total dynamical mass. Details can be found in Table 
\ref{clump_table}. 
In order to obtain a more rigorous estimate for the significance of this discrepancy, we used Monte 
Carlo simulations. Galaxy redshifts were randomly distributed $10^6$ times while keeping galaxy 
positions fixed, destroying any possible correlation between redshift and position. In only 
95 out of the $10^6$ cases do we find a higher mean redshift for the clump around $\mathrm{cD_1}$ 
and a lower one for the other than in the observed data, meaning a $3.9\sigma$ significance. 
In addition, we checked whether the two distributions with $\sigma_{v,1}=879\,\rm{km\;s^{-1}}$ and 
$\sigma_{v,2}=960\,\rm{km\;s^{-1}}$ can be drawn from the same parent distribution with 
$\sigma_{v}=1296\,\rm{km\;s^{-1}}$. We created $10^5$ random realisations of this parent gaussian 
distribution, each containing 21 velocities that were ordered. Every realisation was split into 
two sub-samples, having the same number of overlapping objects in velocity space as the real data, 
shown in Fig. \ref{cl0053_z-spacecorr}. These sub-samples contained 10 and 11 mock galaxies, 
and it was checked whether the difference in their mean velocities was larger than the observed 
$1824\,\rm{km\;s^{-1}}$. We found that in 96.9\% of all cases the observed velocity histogram
could not be drawn from a single gaussian distribution, thus the clumps probably did not yet mix.

We note that the velocity difference between $\mathrm{cD_1}$ and $\mathrm{cD_2}$ themselves, however, 
is $v=866\,\rm{km\;s^{-1}}$ and thus much lower than the difference between the sub-clumps. We conclude
that the 21 measured redshifts for CL0056.03 are probably not representative for the $\sim113$ cluster 
member candidates, as identified by their colours.
\begin{table}[t]
   \caption{\label{clump_table}Properties for the eastern and western clumps around $\mathrm{cD_1}$
respectively $\mathrm{cD_2}$.}
   \scriptsize
   \begin{tabular}{l|lll}
   \hline 
   \noalign{\smallskip}
    Clump & $\langle z\rangle$ & $\langle v\rangle\; [\mathrm{km\;s^{-1}}]$ & 
   $\sigma_v\;[\mathrm{km\;s^{-1}}]$\\
   \noalign{\smallskip}
   \hline 
   \noalign{\smallskip}
   $\mathrm{cD_1}$ & $0.1685\pm0.0029$ & 50521 & 879\\
   $\mathrm{cD_2}$ & $0.1624\pm0.0032$ & 48697 & 960\\
   \noalign{\smallskip}
   \hline 
   \noalign{\smallskip}
   \noalign{\smallskip}
   \end{tabular}
   \small
\end{table}
In the following a simple dynamical model is used in order to check whether the system is 
gravitationally bound or unbound. A linear orbit of the clumps is assumed, i.e. there is no 
shear or rotation component. The system is seen under some inclination angle $\varphi$, with 
$\varphi=0$ if the clumps were aligned along the line of sight. Based on Newtonian dynamics 
the condition of a bound system can be written as $v^2 r \leq 2 G M$, or, taking into account 
the inclination angle $\varphi$, as
\begin{equation}
\label{dyn_limit}
\frac{v^2_{\rm{r}} r_{\rm{p}}}{2 G M}\leq \rm{sin}^2\varphi\,\rm{cos}\,\varphi\;,
\end{equation}
where $v_{\rm{r}}=1824\,\rm{km\,s}^{-1}$ is the radial velocity difference between the clumps and 
$r_{\rm{p}}=300\,\rm{kpc}$ the projected distance on the sky between $\mathrm{cD_1}$ and 
$\mathrm{cD_2}$. $M$ is the total mass of the system. The right hand side of (\ref{dyn_limit}) can 
not get larger than $0.385$ for $\varphi=54^{\circ}$. For $M\leq3.0\times10^{14}\,M_{\odot}$ the 
expression (\ref{dyn_limit}) does not allow bound solutions for the given $v_{\rm{r}}$ and $r_{\rm{p}}$. 
For $M=5.0\times10^{14}\,M_{\odot}$ the probability for a bound system is 49\% (the fraction of allowed 
inclination angles over all inclination angles), and it increases to 70\% for 
$M=1.0\times10^{15}\,M_{\odot}$. Thus, the lensing estimated mass of $M=1.4\times10^{14}\,M_{\odot}$ for 
CL0056.03 is not sufficient for a bound solution. If the velocity difference 
$v_{\rm{r}}=866\,\rm{km\,s}^{-1}$ between the two most luminous galaxies $\mathrm{cD_1}$ and 
$\mathrm{cD_2}$ was representative for CL0056.03, instead, then in 57\% of the cases the system would 
be gravitationally bound.

Our interpretation for the dynamical state of CL0056.03 is thus that the eastern clump is closer to us 
and the western one further away, and that the system is in a pre-collision phase with the two clumps 
approaching each other, since probably no mixing has yet taken place. In addition, there is some 
indication from the imaging data itself, arguing for this scenario: both the eastern and the western 
clump are embedded in large, but well separated haloes, each measuring about 340 kpc. The halo around 
the western clump, in particular, shows tidal features. We take this as an evidence for ongoing tidal 
stripping and merging processes inside the clumps, but probably no encounter has yet taken place between 
them. In the spectra taken by CHM98 no emission lines are found for the cluster members, thus there is 
no sign for nuclear activity or major star formation. The other scenario is that there was already an 
encounter and the two clumps are receding, which would place the western clump closer to the observer.
Based on the small number of spectra the latter state can not be ruled out.

Besides the two sub-clumps there is some evidence for a more extended structure in CL0056.03.
In Fig. \ref{redsequence_total} a filament is seen, extending to about 7\myarcminnodot north 
and 4\myarcminnodot south of the cluster centre, but not lined up with the cluster centre itself. 
The angle between this linear filament and the orientation of CL0056.03 is $\sim40$ degrees.
North of CL0056.03 at ($\alpha=00:55:45.6, \delta=-37:24:46$), however, the small cluster Abell S0102 
is found at a redshift of $z=0.056$. A contamination of the possible filament by member galaxies
of Abell S0102 can not be ruled out. Furthermore, 5\myarcmin2 from the cluster centre and in the
direction of the cluster orientation, an isolated elliptical galaxy with $R=14.4$ and $V-R=0.745$ 
is found, 0.05 mag bluer than the cluster sequence. Its distance to the centre of Abell S0102 is 
6\myarcmin2. Based on the present data it can not be concluded whether this elliptical is isolated 
in space or a physical member at the fringes of one of the two clusters.
\section{Conclusions}
We introduced the Garching-Bonn Deep Survey, a 12 square degree survey for weak lensing investigations.
It is primarily a `virtual' survey, since about 75\% of the data was taken from the ESO archive. Within
a dedicated ASTROVIRTEL program the functionality of the \textit{querator} search engine was substantially
expanded. It now allows an effective filtering of the data with respect to characteristic properties.
One of these archival fields, centreed on NGC 300, showed two concentrations of background galaxies upon
visual inspection of the field in the DSS. The cluster nature of these two concentrations was 
spectroscopically confirmed beforehand by other groups.

Making use of the high quality of the $R$-band exposures, we showed that one of the two clusters 
(CL0056.03) can be detected by its weak gravitational shear signal. The second cluster (CL0056.02) 
was not found in the weak lensing data, probably due to field truncation and insufficient image quality. 
Besides, two other significant and coherent shear patterns were detected. Both of them coincide with 
significant overdensities of red galaxies, for which we gave redshift and mass estimates, based on 
their apparent colour and lensing strength. A third shear detection was seen behind NGC 300 itself, 
but there is no further evidence from background light for a cluster of galaxies at this position. 
Thus, apart from this hidden detection, all $M_{\rm ap}$ peaks equal to or higher than $3\sigma$ in 
filter scales $\geq$ 3\myarcmin2 coincide with overdensities of red galaxies. These $M_{\rm ap}$ 
peaks, however, are at the limit of what can be concluded from this data set. The tidal gravitational 
fields of more massive clusters of galaxies imprint shear fields in the images of background galaxies 
that are easily detected on the $10\sigma$ level and beyond \citep{clowe}.

Based on the high virial mass of CL0056.03, derived from spectroscopic data by CHM98, and the 
comparatively shallow weak lensing signal, a closer analysis of this cluster was performed. A highly 
significant correlation between redshift and the position of galaxies in the cluster is 
found, indicating the existence of two sub-clumps, separated by $1824\,\rm {km\,s^{-1}}$ in velocity 
space. Based on the present data and the derived lensing mass it can not be concluded whether the system is
gravitationally bound. By looking at the projected sky distribution of galaxies with similar properties 
in a $(V-R,R)$ colour-magnitude diagram, we find an extended filament of galaxies inclined to the cluster's 
major axis. Without further spectroscopic data of galaxies in this filament its nature can not
be further clearified. We take the presented properties as evidence that CL0056.03 is a younger 
cluster in formation and has not yet reached its equilibrium state.

CL0056.03 shows a clear and tight red sequence in the $(V-R,R)$ colour-magnitude space, with an 
increasing fraction of bluer galaxies towards larger cluster radii. CL0056.02 also shows a red 
sequence, the results presented for this cluster, however, are more uncertain due to field truncation 
and inferior data quality.
\begin{acknowledgements}
The authors thank Nathalie Fourniol and Benoit Pirenne (ESO archive), for their excellent assistance 
of our frequent and substantial data requests. Furthermore the kind provision of \textit{FLIPS} by 
Jean-Charles Cuillandre is greatly appreciated. MS thanks Matthias Bartelmann, Oliver Czoske, Evanthia 
Hatziminaoglou, Lindsay King, Joan-Marc Miralles and Stella Seitz for providing photometric redshift 
estimates, further discussions and many corrections. The support given by ASTROVIRTEL, a Project 
funded by the European Commission under FP5 Contract No. HPRI-CT-1999-00081, is acknowledged.
This work was supported by the German Ministry for Science and Education (BMBF) through the DLR under 
the project 50 OR 0106, by the German Ministry for Science and Education (BMBF) through DESY under the 
project 05AE2PDA/8, and by the Deutsche Forschungsgemeinschaft (DFG) under the project SCHN 342/3--1.
\end{acknowledgements}

\end{document}